\pdfoutput=1 
\documentclass[aps,pra,letterpaper,notitlepage,longbibliography,floatfix,12pt,reprint]{revtex4-1}

\usepackage{amsmath,amsfonts,amssymb}

\usepackage{float} 

\usepackage{graphicx}
\usepackage[subrefformat=parens, caption=false]{subfig}

\usepackage[margin=0.7in]{geometry}

\usepackage{hyperref}
\hypersetup{colorlinks=true}
\usepackage[all]{hypcap} 

\usepackage{times}

\usepackage{upgreek}

\usepackage{color}

\DeclareMathOperator{\re}{Re}

\newcommand*{\Resize}[2]{\resizebox{#1}{!}{$#2$}}

\newcommand{\figurepanel}[2]{\hyperref[#1]{\ref*{#1}(#2)}}

\allowdisplaybreaks[1] 

\usepackage{filecontents}

\begin{document}
\title{Waveguide QED: Power Spectra and Correlations of Two Photons\\ Scattered Off Multiple Distant Qubits and a Mirror}
\author{Yao-Lung L. Fang}
\author{Harold U. Baranger}
\affiliation{Department of Physics, Duke University, P.O. Box 90305, Durham, North Carolina 27708-0305, USA}
\date{28 April 2015}

\begin{abstract}

We study two-level systems (2LS) coupled at different points to a one-dimensional waveguide in which one end is open and the other is either open (infinite waveguide) or closed by a mirror (semi-infinite). 
Upon injection of two photons (corresponding to weak coherent driving), the resonance fluorescence and photon correlations are shaped by the effective qubit transition frequencies and decay rates, which are substantially modified by interference effects.   
In contrast to the well-known result in an infinite waveguide, photons reflected by a single 2LS coupled to a semi-infinite waveguide are initially bunched, a result that can be simply explained by stimulated emission. As the number of 2LS increases (up to 10 are considered here), rapid oscillations build up in the correlations that persist for a very long time. For instance, when the incoming photons are slightly detuned, the transmitted photons in the infinite waveguide are highly antibunched. On the other hand, upon resonant driving, incoherently reflected photons are mostly distributed within the photonic band gap and several sharp side peaks. These features can be explained by considering the poles of the single particle Green function in the Markovian regime combined with the time delay. Our calculation is not restricted to the Markovian regime, and we obtain several fully non-Markovian results. We show that a single 2LS in a semi-infinite waveguide can \emph{not} be decoupled by placing it at the node of the photonic field, in contrast to recent results in the Markovian regime. Our results illustrate the complexities that ensue when several qubits are strongly coupled to a bus (the waveguide) as might happen in quantum information processing.  


\end{abstract}

\maketitle

\section{Introduction}

The study of multiple photons confined in a one-dimensional (1D) waveguide interacting with local emitters (qubits) \cite{YudsonJETP84, YudsonJETP85, ShenOL05, ShenPRL05, ChangPRL06} has attracted a great deal of attention recently, and is now referred to by the term ``waveguide quantum electrodynamics (QED)''. The 1D geometry greatly limits the possible propagating directions and hence increases the interference effects \cite{ShenPRL05, ChangPRL06} while decreasing the mode volume, which in turn enhances the coupling strength of qubits to the waveguide \cite{BlaisPRA04}. 
The 1D strong-coupling regime, where the light-matter interaction dominates over loss and dephasing, provides an excellent setting in which to 
investigate interesting quantum-optical effects theoretically
\cite{ShenOL05, ShenPRL05, ChangPRL06, ChangNatPhy07, ShenPRL07, ShenPRA07, YudsonPRA08, WitthautNJP10, ZhengPRA10, HouNoriPRA10, LongoPRL10, RephaeliPRA11, RoyPRL11, RoyPRA11, ShiSunPRA11, LongoPRA11, ZhengPRL11, ZhengPRA12, RephaeliPRL12, LongoOptExp12, ZhengPRL13, MoeferdtOL13, LalumierePRA13, PeropadreNJP13, LindkvistNJP14, RoyPRA14, LaaksoPRL14, FratiniPRL14, SanchezPRL14, RedchenkoPRA14, CanevaarXiv15},
observe such effects experimentally,
\cite{BaurPRL09, AstafievSci10, AbdumalikovPRL10, AstafievPRL10, AbdumalikovPRL11, BozyigitNatPhys10, LangPRL11, HoiPRL12, HoiNJP13, HoiPRL13, KoshinoPRL13, vanLooScience13, LangNatPhy13, InomataPRL14, HoiarXiv14}, 
construct building blocks of quantum information processing and quantum computing 
\cite{ShenPRL05,ChangNatPhy07,ZhouPRL08,LongoPRL10,KolchinPRL11,EichlerPRL11, HoiPRL11, LeungPRL12,CiccarelloPRA12,EichlerWallraffPRA12,ZhengOL13,ZhengPRL13a},
and generate qubit-qubit entanglement 
\cite{ChenPRB11, TudelaPRL11, CanoPRB11, GonzalezTudelaPRL13, ZhengPRL13, GonzalezBallestroNJP13, GonzalezBallesteroPRA14}.


A variety of artificial systems have been proposed and realized to implement light-matter interaction in 1D, including superconducting qubits coupled to a microwave transmission line \cite{BaurPRL09, AstafievSci10, AbdumalikovPRL10, AstafievPRL10, AbdumalikovPRL11, BozyigitNatPhys10, LangPRL11, HoiPRL12, HoiNJP13, HoiPRL13, KoshinoPRL13, vanLooScience13, LangNatPhy13, InomataPRL14, HoiarXiv14} or surface acoustic waves \cite{GustafssonSci14}, and semiconductor quantum dots coupled to either a metallic nanostructure \cite{AkimovNat07, VersteeghNatCommun14,AkselrodNatPho14} or a photonic-crystal waveguide \cite{LauchtPRX12, ArcariPRL14, ReithmaierarXiv14}. In addition to artificial atoms, waveguide QED can also be implemented using an ion trap \cite{MeirPRL14}, cold atoms trapped in \cite{BajcsyPRL09} or near \cite{VetschPRL10} an optical fiber, or single molecules doped in an organic crystal filled in a glass capillary \cite{FaezPRL14}. In several of these systems, the coupling of the local emitter to the waveguide dominates by far all other emission or dephasing processes.

The theoretical difficulty of waveguide QED lies in the fact that the waveguide photons are bidirectional while the qubits have arbitrary positions in the waveguide. A position-dependent phase factor is thus introduced even if the coupling strength for each qubit is the same. As a result, while a few photons scattering off one qubit (or multiple co-located qubits) has been extensively studied and exact solutions exist \cite{ShenPRL07, ShenPRA07, YudsonPRA08, WitthautNJP10, ZhengPRA10, LongoPRL10, LongoPRA11, RoyPRL11, RoyPRA11, RephaeliPRA11, ZhengPRA12, KocabasPRA12, RephaeliPRL12, GonzalezBallesteroPRA14, RoyPRA14}, for treating multiple qubits, the Markovian approximation has appeared necessary. Such Markovian multi-qubit, bidirectional waveguide calculations have been pursued recently using several theoretical techniques: a Green function approach \cite{DzsotjanPRB10, DzsotjanPRB11}, the master equation \cite{TudelaPRL11, CanoPRB11, ChangNJP12}, input-output theory \cite{LalumierePRA13, CanevaarXiv15}, and the Lippmann-Schwinger (L-S) equation \cite{ZhengPRL13, FangEPJQT14}. We note, however, one exception: an exact solution was obtained recently for two bidirectional photons scattering off two separated qubits \cite{LaaksoPRL14}. Furthermore, when entering the ultrastrong-coupling regime where the rotating-wave approximation (RWA) fails \cite{BourassaPRA09, NiemczykNatPhys10}, analytical treatments seem impossible, and one has to use numerical methods such as matrix product states \cite{PeropadrePRL13, SanchezPRL14, ZuecoFD14} to explore the many-body physics of photons.

A single qubit in a semi-infinite waveguide is a more complex problem than for an infinite waveguide because of the delay in the reflection from the end and has therefore received considerable attention
\cite{DornerPRA02, DongPRA09, ChenNJP11, PeropadrePRA11, WangPRA12, KoshinoNJP12, TufarelliPRA13, BradfordPRA13, TufarelliPRA14, HoiarXiv14}.
Although an atom placed in front of a mirror in 3D open space has been studied both theoretically \cite{DornerPRA02} and experimentally \cite{EschnerNature01, WilsonPRL03, BushevPRL04, DubinPRL07}, the unconfined light in 3D makes the interference effect weak, and  
one therefore expects a much stronger effect in 1D. 
An exact solution for the wavefunction of the initially excited qubit can be derived by solving the delay-differential equation \cite{DornerPRA02, BradfordPRA13, TufarelliPRA13, TufarelliPRA14}. This solution demonstrates the complicated interference effects caused by the mirror; if the distance to the mirror is large, non-Markovian effects come into play even for a single excitation (qubit or photon) \cite{TufarelliPRA13, BradfordPRA13, TufarelliPRA14}. Under the Markovian approximation, the problem reduces to solving an ordinary differential equation \cite{DornerPRA02, TufarelliPRA14} which is far easier. 
The presence of the boundary in the semi-infinite case (i.e.\ the mirror) causes a modification of qubit frequencies and decay rates by modulating the structure of the photonic environment \cite{KoshinoNJP12, HoiarXiv14}. 
We are not aware of the existence of exact solutions for any cases of multi-photon scattering.

\begin{figure}[tb]
	\centering
	\includegraphics{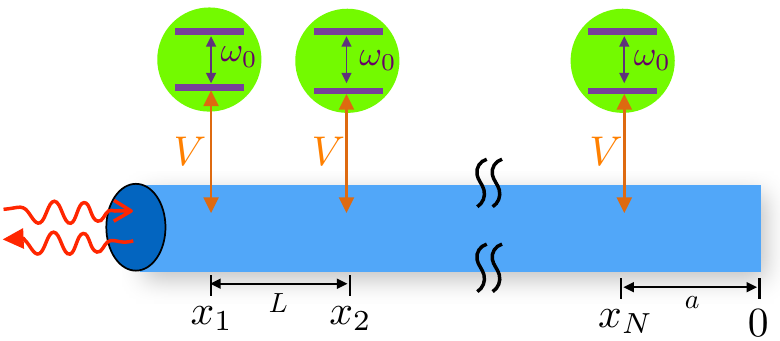}
	\caption{(Color online) Schematic of the waveguide-QED system, in which equally separated, identical 2LS are coupled to a semi-infinite waveguide with one open end and another closed at $x=0$. For an infinite waveguide with two open ends, the 2LS are instead placed symmetrically with respect to $x=0$ to simplify the calculation.
		\label{fig:schematics}}
\end{figure}

In this paper, we consider $N$ identical, equally-separated two-level systems (2LS) strongly coupled to an infinite or semi-infinite waveguide (Fig.~\ref{fig:schematics}, the infinite waveguide has two open ends while the semi-infinite is closed by a perfect reflector on one end). 
Most of the results are obtained using the Markovian approximation, which is checked by a full non-Markovian calculation in a few cases. 
It has been known since the introduction of the Dicke model \cite{DickePR54} that interaction among the multiple 2LS can be induced through their coupling to bosonic modes, leading to sub- and super-readiance.  
In 1D waveguides in particular, recent theoretical \cite{LalumierePRA13} and experimental \cite{vanLooScience13} studies of the power spectrum of two qubits coupled to an infinite waveguide clearly show that the qubit-qubit separation $L$  
modulates the effective resonant frequencies and decay rates, resulting in sub- and super-radiance. 
While it seems natural, then, to explore situations with many qubits, in fact discussion beyond two-qubit systems is limited in the literature \cite{TsoiPRA08, ChangNJP12, RoySciRep13, FangEPJQT14, ZuecoFD14, ZoubiPRA14}.  
Using the Lippmann-Schwinger equation, we show analytically that in the Markovian regime the collective behavior 
is encoded in the simple poles of the Green function. These poles reveal themselves in various measurable quantities such as the transmission spectrum, time delay $\uptau$, power spectrum $S(\omega)$ (resonance fluorescence), and two-photon correlation functions $g_2(t)$ (second-order coherence). The Markovian approximation reduces the number of poles from infinity to $N$ \cite{ZhengPRL13} and so renders the problem tractable. Throughout the paper, we highlight a number of common features of our results, such as rapidly oscillating two-photon correlations that persist for a long time, and the concentration of the reflected fluorescence within the photonic band gap along with sharp side peaks. 

We point out an intriguing difference between the infinite and semi-infinite waveguides: while a single qubit coupled to the former can only reflect one photon at a time, giving rise to initial anti-bunching \cite{ShenPRA07, ChangNatPhy07, GardinerQN00},
when coupled to the latter it instead bunches the reflected photons. This can be explained simply by the stimulated emission. 
Another effect in a semi-infinite waveguide is the possibility of decoupling the waveguide from the 2LS by placing it at a node of the single-photon wavefunction when the qubit-mirror separation is small, as studied theoretically \cite{KoshinoNJP12} and experimentally verified using superconducting qubits \cite{HoiarXiv14}. We show that if the distance is large, however, the non-Markovian effects that come into play destroy this decoupling: our numerical non-Markovian calculation in the two excitation sector shows that the 2LS remains coupled to the waveguide because of oscillating nontrivial correlations. These two-photon features, to the best of our knowledge not addressed by previous 1D studies \cite{DongPRA09, ChenNJP11, PeropadrePRA11, WangPRA12, KoshinoNJP12, TufarelliPRA13, BradfordPRA13, TufarelliPRA14}
which mainly concern single-excitation properties, should be readily measurable using existing experimental technology.


The rest of this paper is organized as follows: We first devote Sec.~\ref{Sec:infinite-S(w)} to discussing the power spectrum of two photons scattering off multiple distant qubits coupled to an infinite waveguide using the L-S equation. Since the power spectrum is a ``first-order'' quantity, one expects it to be easier to calculate and measure. In Sec.~\ref{Sec:infinite-g2} we then move on to results for the second-order photon correlation $g_2(t)$. To explain the long-time behavior of $g_2$, we introduce the concept of time delay in Sec.~\ref{Sec:timedelay}. In Sec.~\ref{Sec:semi-infinite-S(w)} and \ref{Sec:semi-infinite-g2}, we turn to the discussion of power spectra and correlations for the semi-infinite waveguide. Some technical details are left for the appendices, including the details of the L-S equation for both infinite and semi-infinite waveguides, the demonstration of the equivalence between the L-S equation and input-output theory at weak coherent driving, and finally the two-photon transmission and reflection probabilities calculated using the L-S equation.


\section{Multiple Qubits in an Infinite Waveguide: Power Spectra}
\label{Sec:infinite-S(w)}

Our starting point is the standard Hamiltonian used in waveguide QED
\cite{ShenPRL05, ChangNatPhy07}, consisting of a one-dimensional bosonic field coupled to discrete 2LS. After making the rotating-wave approximation (RWA) and extending the limits of the momentum integrals to infinity, one finds that the Hamiltonian in real space is (taking $\hbar=c=1$) 
\begin{align}
&H=H_\text{qubit}
-i\int\limits_{-\infty}^{\infty} dx \left[a^\dagger_\text{R}(x)\frac{d}{dx}a_\text{R}(x)-a^\dagger_\text{L}(x)\frac{d}{dx}a_\text{L}(x)\right]\nonumber\\
&+\sum_{i=1}^N\sum_{\alpha=\text{L,R}} \!V\! \int\limits_{-\infty}^{\infty} \!dx\; \delta(x-x_i)\left[a^\dagger_\alpha(x)\sigma_{i-}+\sigma_{i+} a_\alpha(x)\right],\label{eq:Hamiltonian}
\end{align}
where $H_\text{qubit}=\omega_0\sum_{i=1}^N\sigma_{i+}\sigma_{i-}$, $\sigma_{i\pm}$ denotes the Pauli raising (lowering) operator of the \textit{i}-th qubit with frequency $\omega_0$ and position $x_i$, 
$a_\text{R,L}$ denotes the annihilation operator of right- (left-) going photons, and $V$ is the coupling strength between the qubit and the photons. The decay rate for each qubit (to the waveguide) is $\Gamma \equiv 2V^2$. Throughout this paper we focus on the lossless limit, but loss could be simply introduced by 
tracing out an auxiliary waveguide or  
modifying the S-matrix elements \cite{Carmichael93, KoshinoNJP12, RephaeliPR13}.

We calculate physical quantities by using the wavefunction $|\psi_2\rangle\equiv|\psi_2(k_1,k_2)\rangle_\text{RR}$ of two incoming right-going photons (RR) with momenta $k_1$ and $k_2$. Throughout this work we focus solely on identical incident photons: $k_1=k_2=E/2$ where $E$ is the total input energy. The first physical quantity we consider is the power spectrum or resonance fluorescence, which is simply the Fourier transform of the first-order coherence, 
\begin{equation}
S_\alpha(\omega)=\int dt\,e^{-i\omega t}
\langle\psi_2| a_\alpha^\dagger(x_0) a_\alpha(x_0+t) |\psi_2\rangle,
\end{equation}
where $x_0$ denotes the detector position (or equivalently, time) far away from the scattering region. $S_\alpha(\omega)$ is simply the spectral decomposition of the photons in the wavefunction $|\psi_2\rangle$. Since the Hamiltonian $H$ preserves the  number of excitations (photon plus qubit), one can insert a one-particle identity operator, $\mathcal{I}_1=\sum_\alpha\int dk|\phi_1(k)\rangle_\alpha\langle\phi_1(k)|$, between the photon operators:
\begin{align}
S_\alpha(\omega)&=\sum_{\alpha'=\text{R,L}}\int dk\int dt\,e^{-i\omega t} 
\nonumber \\
& \times \langle\psi_2|a_\alpha^\dagger(x_0)|\phi_1(k)\rangle_{\alpha'}\langle\phi_1(k)|a_\alpha(x_0+t)|\psi_2\rangle ,
\label{eq:power spectrum in scattering theory}
\end{align}
where $|\phi_1(k)\rangle_\alpha$ is the single-particle scattering eigenstate satisfying $H\,|\phi_1(k)\rangle_\alpha=k|\phi_1(k)\rangle_\alpha$ with the incoming wave traveling in the $\alpha=$ L or R  direction. The power spectrum follows by computing the matrix elements $_{\alpha'}\langle\phi_1(k)|a_\alpha(x_0+t)|\psi_2\rangle$. 

The two-photon wavefunction $|\psi_2\rangle$ is obtained via the Lippmann-Schwinger equation following the procedure in Refs.~\cite{ZhengPRL13, FangEPJQT14}. 
The building blocks are the single-particle eigenstates $|\phi_1\rangle$, the two-particle states $|\phi_2\rangle$ formed from the direct product of two $|\phi_1\rangle$, and the corresponding retarded Green function $G^R(E)$. In fact, $|\psi_2\rangle$ can be written as [see Eq.~(20) in Ref.~\onlinecite{FangEPJQT14}] 
\begin{align}
&\lefteqn{| \psi_2(k_1, k_2)\rangle_{\alpha_1,\alpha_2}
 =| \phi_2(k_1, k_2)\rangle_{\alpha_1,\alpha_2} }  \nonumber \\
& -\sum_{i,j=1}^N
G^R(E)|d_i d_i \rangle \left(G^{-1}\right)_{ij}\langle d_j d_j|\phi_2(k_1, k_2)\rangle_{\alpha_1,\alpha_2}. 
\label{eq:two-photon wavefunction}
\end{align}
The second term contains all the nonlinearity and is often referred to as the two-photon ``bound state'' \cite{ShenPRA07, ZhengPRA10, RoyPRA11}.
One can evaluate the desired matrix elements by inserting two-particle identity operators $\mathcal{I}_2$ in the second term of Eq.~\eqref{eq:two-photon wavefunction} and performing the double momentum integral thereby introduced [see Eq.~(23) in Ref.~\onlinecite{FangEPJQT14}].
This calculation is exact. For more than one qubit, making the Markovian approximation allows the integration to be done analytically. In this context, the Markovian approximation consists in replacing all factors of $\exp(ikL)$ that occur in the Green functions in Eq.~\eqref{eq:two-photon wavefunction} by $\exp(ik_0L)$, where $k_0=\omega_0/c$ is the wavevector associated with $\omega_0$ and $L=x_{i+1}-x_i$ is the qubit-qubit separation \cite{FangEPJQT14}. In practice, we use a slightly modified expression for the power spectrum, 
\begin{subequations}
\label{eq:power_spectrum_in_practice}
\begin{align}
S_\text{R}(\omega)=&2\,\text{Re}\sum_{\alpha'}\int dk\int_0^\infty dt\,e^{-i\omega t} \\
& \times \langle\psi_2|a_\text{R}^\dagger(x_0)|\phi_1(k)\rangle_{\alpha'}\langle\phi_1(k)|a_\text{R}(x_0+t)|\psi_2\rangle, \nonumber \\
S_\text{L}(\omega)=&2\,\text{Re}\sum_{\alpha'}\int dk\int_{-\infty}^0 dt\,e^{i\omega t}  \\
& \times \langle\psi_2|a_\text{L}^\dagger(x_0)|\phi_1(k)\rangle_{\alpha'}\langle\phi_1(k)|a_\text{L}(x_0-t)|\psi_2\rangle. \nonumber
\end{align}
\end{subequations}
The calculation of the matrix elements is given in Appendix \ref{appendix:matrix elements}.

\begin{figure}[tb]
	\centering
	\includegraphics[scale=0.95]{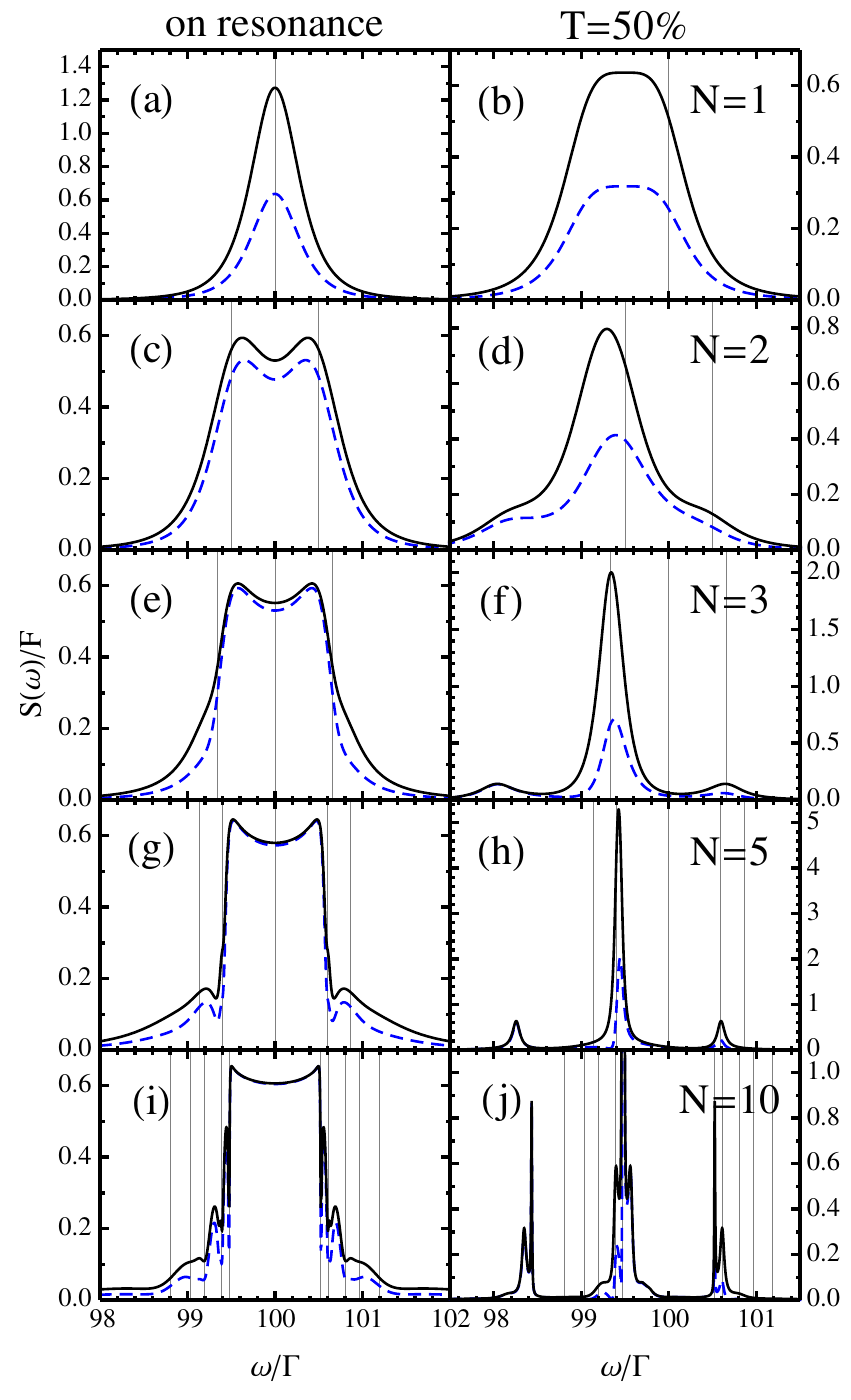}
	\caption{(Color online) Normalized power spectra (resonance fluorescence) of multiple qubits (from top to bottom: $N=1, 2, 3, 5, 10$) coupled to the infinite waveguide with $k_0L=\pi/2$ (separation $L=\lambda_0/4$). For the first column, the incoming photons are on resonance, $E/2=\omega_0=100\Gamma$; for the second column the frequency is chosen such that the single photon transmission is 50\% in each case. The total fluorescence (black solid line) is broken down into the reflected (blue dashed) and transmitted (red dotted) components. The vertical lines indicate the real part of the poles. 
	For $N=10$ in the off-resonant case, the height of the central peak goes up to $\sim60$, which is not shown for better visibility.
	The frequencies used in the second column are $E/2\Gamma=\{99.5, 99.29, 99.34, 99.43, 99.48\}$ (from top to bottom) \cite{precision}. 
		\label{fig:PSD_Piover2_infinite}}
\end{figure}

\begin{figure*}[!htbp]
	\centering
	\includegraphics[scale=0.8]{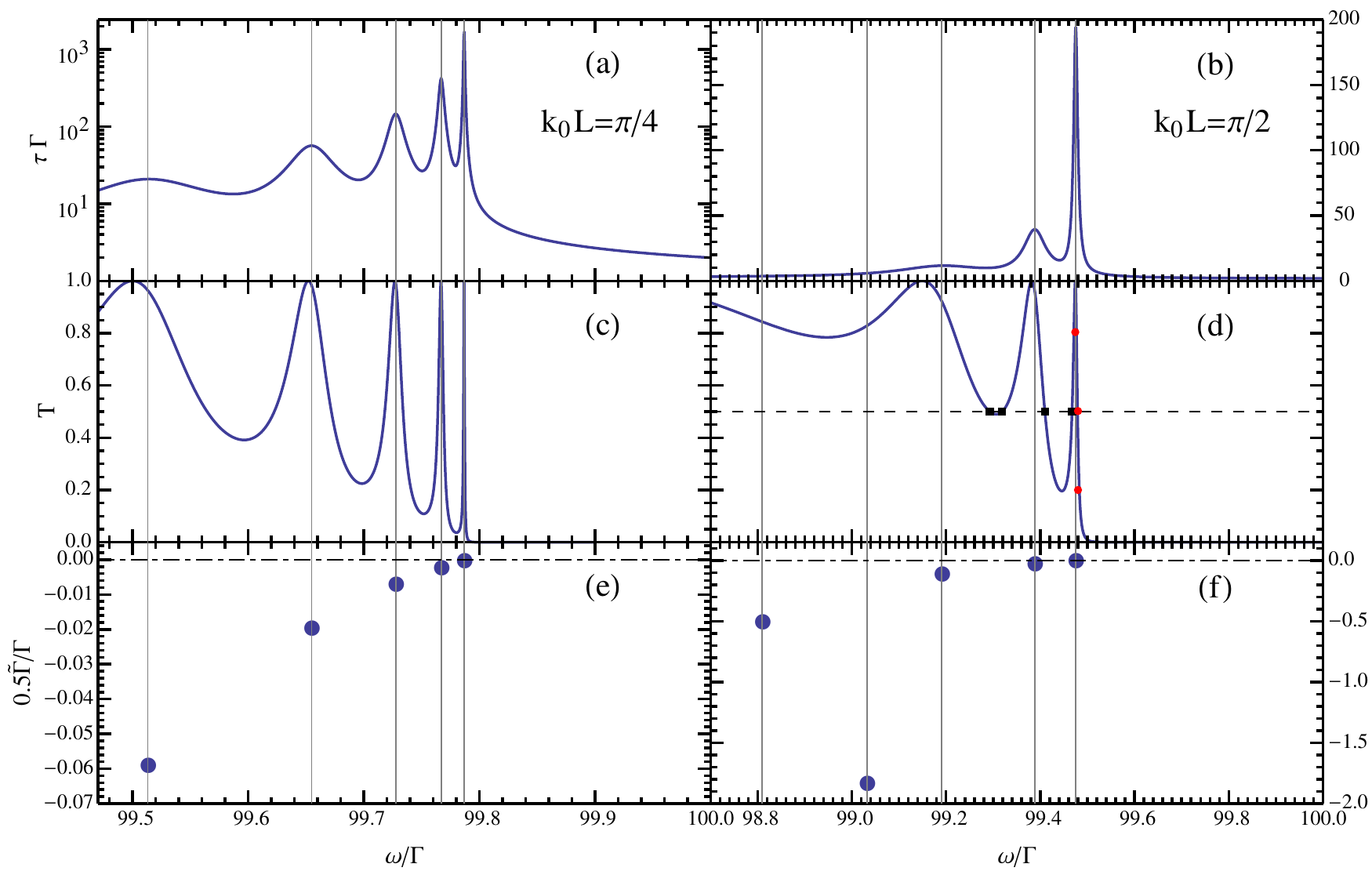}
	\caption{(Color online) Time delay $\uptau$ (top), single-photon transmission spectrum $\text{T}=|t(k)|^2$ (middle), and poles of the transmission amplitude $t(k)$ (bottom) as a function of frequency. The system consists of 10 qubits coupled to an infinite waveguide with $k_0L=\pi/4$ (left column) and $k_0L=\pi/2$ (right column). For the sake of clarity we show only the red-detuned side; for $k_0L=\pi/2$ the poles are symmetric with respect to the qubit frequency $\omega_0=100\Gamma$, while for $k_0L=\pi/4$ five poles are not shown. The vertical lines indicate the real parts of the poles. In panel (d) the black squares give the incident frequencies used in Fig.~\ref{fig:g2_Piover2_T50_all_infinite}, the red dots give those used in Fig.~\ref{fig:g2_Piover2_long_time_infinite}, and the dashed line labels $\text{T}=50\%$. The dashed-dotted lines in panels (e) and (f) label the origin ($\tilde{\Gamma}=0$).
		\label{fig:time_delay_transmission_pole_10qubit_infinite}}
\end{figure*}

After combining all pieces together, the resulting power spectrum can be divided into two parts,
\begin{equation}
S_\alpha(\omega)=S^\text{coherent}_\alpha(\omega)+S^\text{incoherent}_\alpha(\omega)\label{coherent+incoherent},
\end{equation}
where the former contains terms proportional to $\delta(0)\delta(\omega-E/2)$ because delta-normalized plane waves are used, and the latter is zero in the absence of the two-photon bound state and remains finite (for more discussion see Appendix~\ref{appendix:Two-photon Transmission Probability}). Since the total incoherent/inelastic power spectrum is the sum of right- and left-going incoherent power spectra,
\begin{equation}
S^\text{incoherent}(\omega)=S_\text{R}^\text{incoherent}(\omega)+S_\text{L}^\text{incoherent}(\omega),
\end{equation}
one can normalize $S^\text{incoherent}_\alpha(\omega)$ in terms of the incoherently scattered photon ``flux''
\begin{equation}
F^\text{incoherent}=\int d\omega\,S^\text{incoherent}(\omega)\label{Fincoherent}.
\end{equation}
In this paper we omit the superscript ``incoherent'' for simplicity and focus on the inelastic power spectra normalized by $F$ so that shapes and features can be readily compared.
Furthermore, both on- and off-resonance cases are studied. In the off-resonant case, for a fair comparison of systems in which the number of qubits is different, we choose the incident frequency such that (i)~the single-photon transmission probability T is 50\% in each case and (ii)~it is the closest such frequency to the bare qubit frequency $\omega_0$ (see discussion in Ref.~\cite{FangEPJQT14}). Because we mostly focus on cases with small separation, $k_0L\leq\pi/2$ (so $L\leq\lambda_0/4$ with the wavelength $\lambda_0=2\pi c/\omega_0$), this choice leads to red-detuned incident frequencies, as will become clear in the following discussion. 

In interpreting the results, it will be useful to refer to the poles of the system, by which we mean the zeros of the denominator of the single-photon transmission or reflection amplitudes $t(k)$ or $r(k)$ 
\footnote{In contrast to Ref.~\cite{TsoiPRA08}, we find that it is not always true that the denominator of $e_i(k)$ gives $N$ poles for all $i=1, \cdots, N$. For instance, with $N=5$ the wavefunction of the central qubit $e_3(k)$ has only 3 poles. Therefore, it is safer to look at the transmission or reflection amplitudes, $t(k)$ or $r(k)$}. 
Denote the poles by $\tilde{z}_i=\tilde{\omega}_i-i\tilde{\Gamma}_i/2$ with $i=1, 2, \cdots, N$ (the factor of one half is in accordance with the definition of the decay rate $\Gamma$); then, the denominator of $t(k)$ and $r(k)$ can be written as a polynomial of degree $N$,
$(k-\tilde{z}_1)(k-\tilde{z}_2)\cdots(k-\tilde{z}_N)$.
We will see that this indeed gives us the effective qubit frequency and decay rate, as implied by the notation. 
In special cases the poles may be symmetrically arranged with respect to the $\omega=\omega_0$ line. This happens when $k_0L=\pi/2$ because a wavefunction that is even about the middle of the interval between two adjacent qubits has the same magnitude at the site of those qubits as a wavefunction that is odd. For other values of $k_0L$, the amplitude in these two cases is different, leading to an asymmetrical situation in which there are superradiant and subradiant modes.

The power spectra with qubit-qubit separation $k_0L=\pi/2$ are presented in Fig.~\ref{fig:PSD_Piover2_infinite}. The result for the $N=1$ case can be derived exactly and is given in Appendix A [Eq.~\eqref{eq:single qubit S(w) infinite waveguide}]. In general, when the system is driven resonantly ($E/2=\omega_0$) and the pole distribution is symmetric with respect to $\omega_0$ (see Fig.~\ref{fig:time_delay_transmission_pole_10qubit_infinite} for a representative plot), both the transmitted and reflected power spectra are symmetric.
In contrast, when the system is driven off-resonantly, neither the transmission nor the reflection fluorescence is symmetric. However, the total fluorescence (transmission + reflection) is still symmetric with respect to the incident frequency, indicating the conservation of energy and serving as a validity check on our calculation. In addition, thanks to the symmetric pole distribution for $k_0L=\pi/2$, the fluorescence when the incident photons are blue-detuned can be simply obtained by mirroring the red-detuned fluorescence with respect to $\omega_0$ (data not shown).

With regard to the dependence on the number of qubits, the main feature of the power spectra with resonant driving is that the photonic band gap develops, 
resulting in the decrease (increase) of transmission (reflection) fluorescence within the photonic band gap. In addition, many sharp side peaks appear around the photonic band gap, whose positions are roughly labeled by the real parts of the poles $\{\tilde{\omega}_i\}$. Since in general for large $N$ the poles closer to $\omega_0$ have smaller decay rates, we find that both the peak position and peak width could be explained by inspecting the poles' real parts $\{\tilde{\omega}_i\}$ and imaginary parts $\{\tilde{\Gamma}_i\}$, respectively. Finally, our two-qubit $S(\omega)$ agrees with the result obtained from input-output theory with weak coherent driving 
\footnote{See Ref.~\cite{LalumierePRA13}; in making a comparison, note that our definition of total fluorescence is different from theirs (private communication with K.~Lalumi\`ere and A.~Blais).}, 
revealing the fact that two-photon scattering is the dominant process for weak driving. Further discussion is deferred to Appendix \ref{appendix:input-output theory}.

Furthermore, with slightly off-resonant driving the power spectra become sharply peaked. 
These sharp peaks reveal the existence of sub-radiant poles (with $\tilde{\Gamma}<\Gamma$). Taking the $N=10$ case as example [Fig.~\figurepanel{fig:PSD_Piover2_infinite}{j}], since the driving frequency is very close to the pole with the smallest $\tilde{\Gamma}$, that pole is highly excited and gives rise to the central peak with a very small width. The smaller peak on the right has the same $\tilde{\Gamma}$ as the central peak and hence is visible too. Energy conservation then requires the smaller peak on the left to pop up as well. 
Thus, the fact that the poles largely determine the peak position and width is more transparent in the off-resonant cases, at least for those sub-radiant poles. We note that in any case transmission fluorescence is suppressed within the photonic band gap as expected.

\begin{figure}[tb]
	\centering
	\includegraphics[scale=0.95]{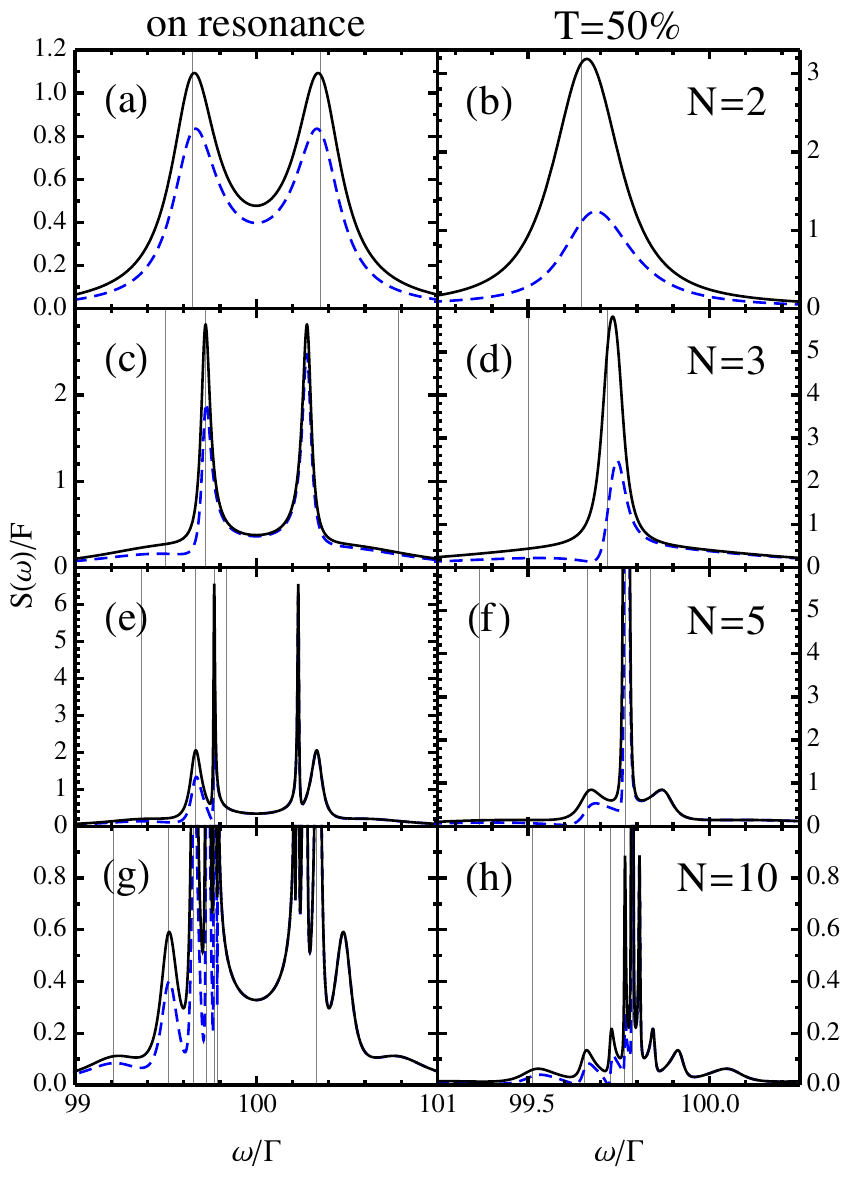}
	\caption{(Color online) Normalized power spectra (resonance fluorescence) of multiple qubits (from top to bottom: $N=2, 3, 5, 10$) coupled to the infinite waveguide with separation $k_0L=\pi/4$ ($L=\lambda_0/8$). The incoming photon frequency is $E/2=\omega_0=100\Gamma$ for the first column (on resonance) and is chosen such that $\text{T}=50$\% for the second column. The total fluorescence (black solid line) is broken down into the reflected (blue dashed) and transmitted (red dotted) components. The vertical lines indicate the real part of the poles. The frequencies used in the second column are $E/2\Gamma=\{99.66, 99.73, 99.77, 99.78\}$ (from top to bottom) \cite{precision}.
	\label{fig:PSD_Piover4_infinite}}
\end{figure}

Next, we consider a smaller separation between the qubits, $k_0L=\pi/4$, see Fig.~\ref{fig:PSD_Piover4_infinite}. For the resonant cases, the main difference from the previous geometry $k_0L=\pi/2$ is that the large reflection fluorescence around $\omega_0$ is reduced. Although it is still true that the reflection fluorescence is higher than the transmission fluorescence, the shape of the photonic band gap is different (red-detuned side is sharper than the blue-detuned side, connected to the asymmetric distribution of poles) making distinct peaks around $\{\tilde{\omega}_i\}$ more visible. Energy conservation implies, as before, that the total power spectrum is symmetric with respect to $\omega_0$. The off-resonant sequence shows similar behavior to the $k_0L=\pi/2$ case, with one minor difference that the blue-detuned power spectra are different from the red-detuned spectra. In general the blue-detuned ones are much smoother because poles on the blue-detuned side ($\tilde{\omega}_i>\omega_0$) have larger decay rate $\tilde{\Gamma}$. Due to limited space we do not show them here.

\section{Multiple Qubits in an Infinite Waveguide: Photon Correlations}
\label{Sec:infinite-g2}

\begin{figure*}[tb]
	\centering
	\includegraphics[scale=0.95]{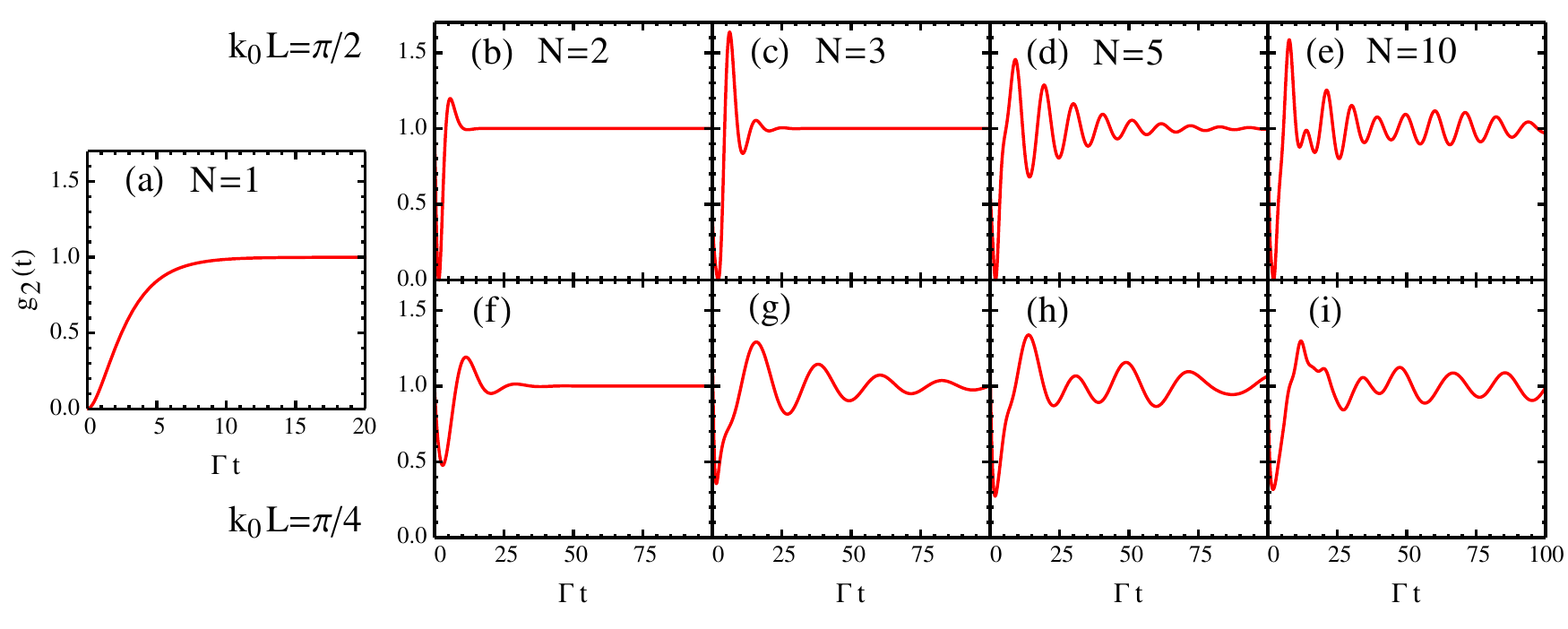}
	\caption{(Color online) Reflection $g_2$ of multiple qubits (from left to right: $N=1, 2, 3, 5, 10$) coupled to an infinite waveguide with separation $k_0L=\pi/2$ (first row) and $\pi/4$ (second row). The two incoming photons are on resonance ($E/2=\omega_0=100\Gamma$). This comparison shows that the limitation $g_2(0)=0$ for single 2LS is removed by adding more 2LS, and that a chain of few 2LS can cause long-time beating.
		\label{fig:g2_Piover2_Piover4_on_resonance_infinite}}
\end{figure*}

We now use the two-photon wavefunction $|\psi_2\rangle$ to calculate the second-order photon correlation function $g_2(t)$ (second order coherence),
\begin{equation}
g_2(t) \equiv \frac{\langle\psi_2|a^\dagger_\alpha(x_0)a^\dagger_{\alpha}(x_0+t)a_{\alpha}(x_0+t)a_\alpha(x_0)|\psi_2\rangle}{|\langle\psi_2|a^\dagger_\alpha(x_0)a_\alpha(x_0)|\psi_2\rangle|^2}.
\label{eq:def_g2}
\end{equation}
Since we are working in the two-photon sector, the numerator implies that $g_2$ is proportional to $|\langle x_0, x_0+t|\psi_2\rangle|^2$ \cite{FangEPJQT14}. 
We first show the cases with resonant driving and $k_0L=\pi/2$ or $\pi/4$ in Fig.~\ref{fig:g2_Piover2_Piover4_on_resonance_infinite}. Although the emergence of the two-photon bound state increases the probability for two photons to be transmitted \cite{ZhengPRA10}, the L-S formalism in which an incoming plane-wave state is used gives an infinitesimally small correction from the two-photon transmission (see Appendix~\ref{appendix:Two-photon Transmission Probability} for details). Therefore, for simplicity we can ignore the transmission $g_2$ and focus on reflection $g_2$ for the resonant cases.

It is well-known that a single 2LS cannot emit two photons at once because it can absorb only one photon at a time, so $g_2(0)=0$ in the reflection channel for $N=1$ \cite{ShenPRA07, ChangNatPhy07, GardinerQN00}. In contrast, we can see from Fig.~\ref{fig:g2_Piover2_Piover4_on_resonance_infinite} that adding more 2LS removes this limitation and allows $g_2(0)$ to be non-zero. The reason is that when one photon is trapped within the first 2LS, the other has a small chance to propagate to and be reflected by the next 2LS, which in turn can cause the stimulated emission of the first photon. Thus, the probability of two photons coming out together is not fully suppressed, a scenario that is even more dramatic for the semi-infinite waveguide treated below.

Secondly, note how oscillations build up and persist for a long time as $N$ increases. We find that the frequency of long-time oscillations matches the difference between the incoming photon frequency $E/2=\omega_0$ (the resonant frequency) and $\tilde{\omega}_i$, the real part of the pole with the smallest decay rate $\tilde{\Gamma}_i$. Since the pole with the smallest decay rate occurs near the edge of the photonic band gap while the resonant frequency is near the middle of the gap, this low frequency scale should be $\omega \sim 0.5\Gamma$ which is indeed what we observe. This makes sense since poles with larger decay rates have much less contribution to $g_2$ at long time. In other words, we see the beating between the most sub-radiant pole and the driving frequency.

\begin{figure}[tb]
	\centering
	\includegraphics[scale=0.9]{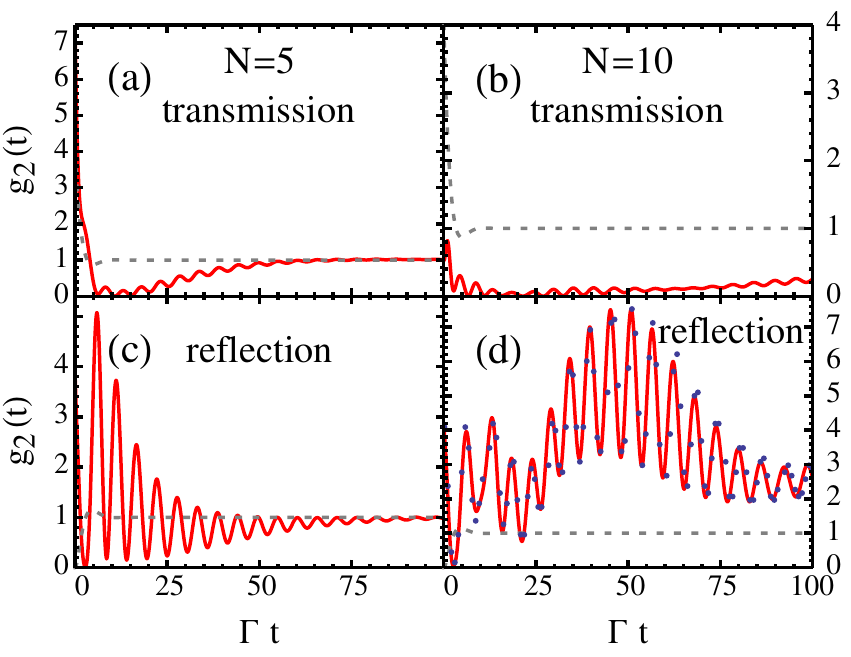}
	\caption{(Color online) In the off-resonant case, $g_2$ of multiple qubits (left: $N=5$; right: $N=10$) coupled to an infinite waveguide with separation $k_0L=\pi/2$. The first row is for two transmitted photons and the second for two reflected ones. The gray, dashed line is the $N=1$ result serving as a reference. The solid curve is calculated using the Markovian approximation while the full non-Markovian result is given by the dots. The frequency of the incoming photons is chosen such that $\text{T}=50$\%, and the qubit frequency is $\omega_0=100\Gamma$.
		\label{fig:g2_Piover2_T50_infinite}}
\end{figure}

We next discuss $g_2$ in the off-resonant cases. Because the $N=2$ and $3$ cases have been discussed in Ref.~\cite{FangEPJQT14}, here we only present results for $N=5$ and $10$. For $k_0L=\pi/2$, they are shown in Fig.~\ref{fig:g2_Piover2_T50_infinite}. It is known that, for $k_0L=\pi/2$ and $N=3$, the transmission $g_2$ has a large initial bunching ($g_2>1$), while the reflection $g_2$ oscillates around the uncorrelated value 1~\cite{FangEPJQT14}. It is striking that as $N$ increases, the transmission correlations show antibunching ($g_2<1$) over a very long time, and the initial bunching is even diminished in the $N=10$ case. The reflection $g_2$ continues to show a great deal of oscillation but in addition becomes highly bunched ($g_2>1$). The oscillation can be explained, as in the resonant case, by the beating between the most sub-radiant poles and the driving. 

We checked these results that use the Markovian approximation against fully non-Markovian numerical results in a few cases. One of them is shown in Fig.~\figurepanel{fig:g2_Piover2_T50_infinite}{d}. The agreement between the two calculations (compare dots and solid line) is very good, showing that the Markovian approximation is reasonable for a qubit chain of moderate size. 

\begin{figure}[tb]
	\centering
	\includegraphics[scale=0.9]{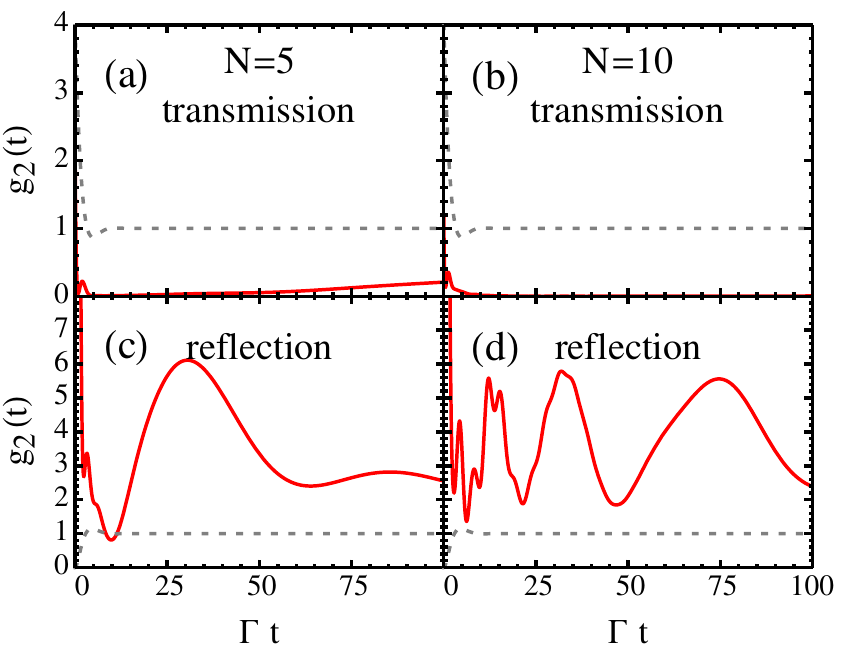}
	\caption{(Color online) In the off-resonant case, $g_2$ of multiple qubits (left: $N=5$; right: $N=10$) coupled to an infinite waveguide with separation $k_0L=\pi/4$. The first row is for two transmitted photons and the second for two reflected ones. The gray, dashed line is the $N=1$ result serving as a reference. The frequency of the incoming photons is chosen such that T=50\%, and the qubit frequency is $\omega_0=100\Gamma$.
		\label{fig:g2_Piover4_T50_infinite}}
\end{figure}

For $k_0L=\pi/4$ and off-resonant photons, the $g_2$ correlation is shown in Fig.~\ref{fig:g2_Piover4_T50_infinite}. As in the $N=3$ case~\cite{FangEPJQT14}, there is sharp initial bunching for \emph{both} reflection and transmission. At non-zero $t$, the reflection $g_2$ shows bunching with irregular oscillation while the transmission photons become strongly antibunched for a long time with little oscillation visible. The reason that $g_2$ of $k_0L=\pi/4$ is very different from $k_0L=\pi/2$ can be attributed to the highly asymmetric pole distribution. Take the $N=10$ case as an example for which the poles are shown in Fig.~\ref{fig:time_delay_transmission_pole_10qubit_infinite}: there are two very close, sub-radiant poles that can contribute to the beating, and the beating frequency is small (one order of magnitude smaller than the $\pi/2$ case) since we choose the incoming frequency to be red-detuned. The complicated interference effects result in highly nontrivial oscillations. We note that the oscillation is gradually washed out beyond $\Gamma t=100$ (data not shown).

\begin{figure*}[tb]
	\centering
	\includegraphics[scale=0.9]{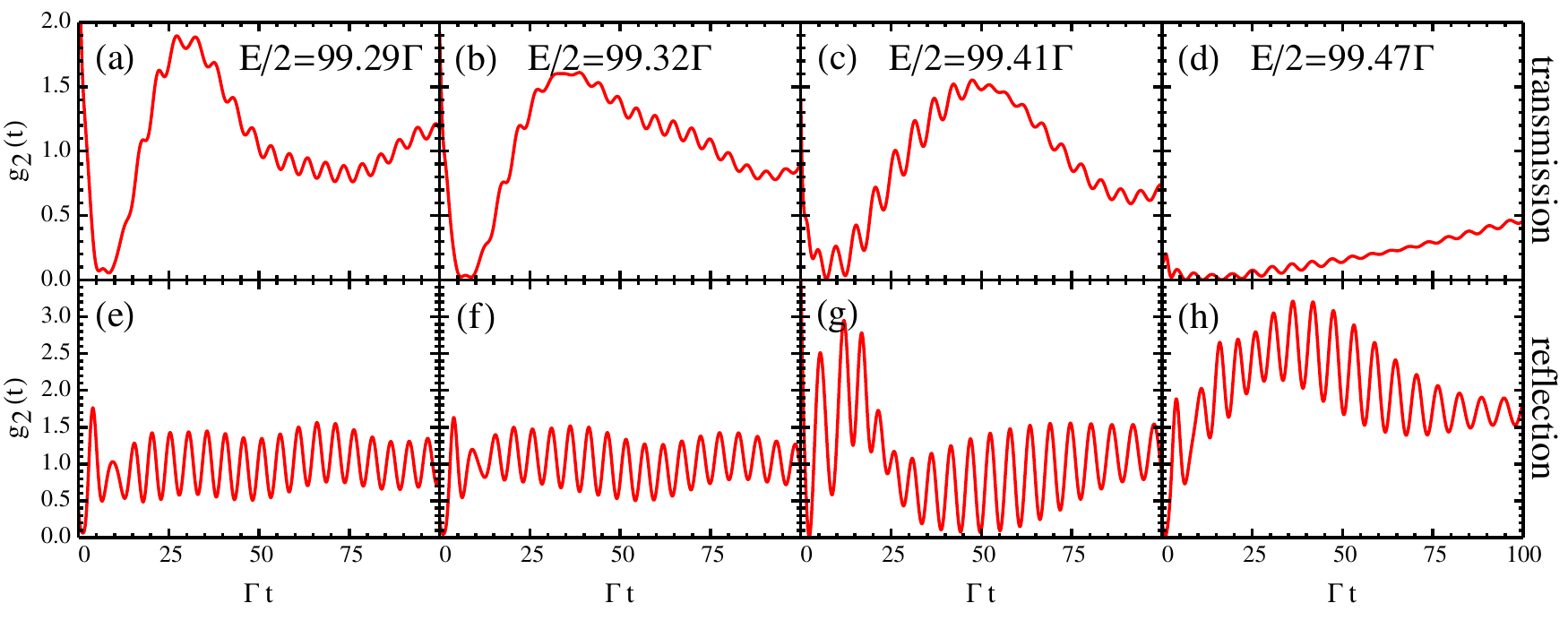}
	\caption{(Color online) $g_2$ of 10 qubits coupled to an infinite waveguide with separation $k_0L=\pi/2$. The first (second) row is for two transmitted (reflected) photons. The driving frequencies are chosen such that $\text{T}=50\%$ and are labeled as black squares on the single-photon transmission spectrum in Fig.~\figurepanel{fig:time_delay_transmission_pole_10qubit_infinite}{d}. The qubit frequency is $\omega_0=100\Gamma$.
		\label{fig:g2_Piover2_T50_all_infinite}}
\end{figure*}

From the above results, $g_2$ is clearly very sensitive to the qubit-qubit separation $L$ and the driving frequency (frequency of incoming photons). One may notice, however, that the resonant cases with $\pi/2$ and $\pi/4$ (Fig.~\ref{fig:g2_Piover2_Piover4_on_resonance_infinite}) are somewhat more similar to each other and distinct from the off-resonant cases. Upon inspecting the polynomials giving rise to the poles for various system configurations, we find empirically that there is a general relation between the $N$ poles,
\begin{equation}
\frac{1}{N}\sum_{i=1}^N \tilde{z}_i =\omega_0-\frac{i\Gamma}{2};
\end{equation}
that is, the average or ``center of mass'' of the poles coincides with the 2LS frequency and decay rate. This relation is \emph{independent} of $L$ and therefore provides a hand-waving explanation: upon resonant driving ($E/2=\omega_0$), the incoming photon frequency always matches the typical, average frequency of the excitations, leading to considerable absorption and reemission and and hence correlation.  

For the off-resonant case, we have chosen a particular value of the frequency for which T, the transmission, is 50\%. There are, potentially, many such frequencies for a given system, and so we turn to comparing the behavior at these different points. As an example, we take the $N=10$, $k_0L=\pi/2$ case. The chosen frequencies are labeled in Fig.~\figurepanel{fig:time_delay_transmission_pole_10qubit_infinite}{d}; note that they are progressively further away from the resonance $\omega_0$. The result is shown in Fig.~\ref{fig:g2_Piover2_T50_all_infinite}. It is clear that the behavior is indeed somewhat different for the five chosen frequencies. We first note that they all oscillate at roughly the same frequency due to the beating with the most sub-radiant poles. 
Secondly, the long-time structure of $g_2$ increases as the detuning becomes smaller, meaning that when driving very close to the frequency of the most sub-radiant pole [about $99.48\Gamma$ in this case; see Fig.~\figurepanel{fig:time_delay_transmission_pole_10qubit_infinite}{f}], a large time-scale sets in, leading to the long-time structure in $g_2$. As we shall see below, this is attributed to the large time delay associated with the most sub-radiant pole. 

\begin{figure}[tb]
	\centering
	\includegraphics[scale=0.9]{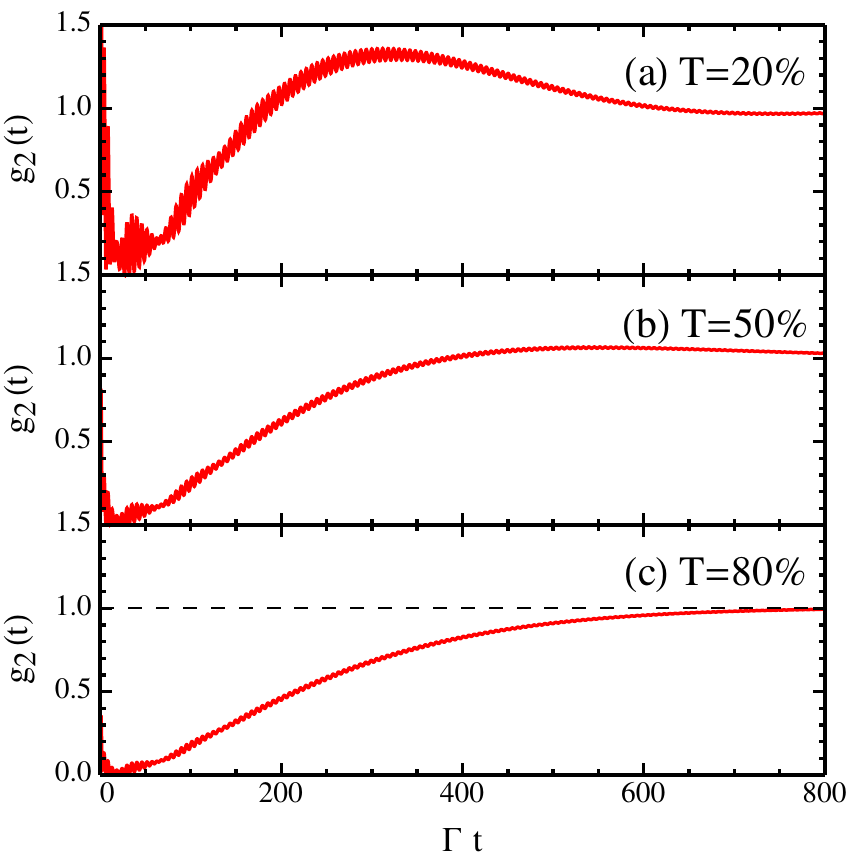}
	\caption{(Color online) Transmission $g_2$ of 10 qubits in an infinite waveguide with separation $k_0L=\pi/2$. The driving frequencies are chosen such that the single particle transmission is (a) 20\%, (b) 50\%, and (c) 80\% [labeled by red dots in Fig.~\figurepanel{fig:time_delay_transmission_pole_10qubit_infinite}{d}], close to the most sub-radiant pole. The qubit frequency is $\omega_0=100\Gamma$.
		\label{fig:g2_Piover2_long_time_infinite}}
\end{figure}

In fact, if one drives very close to the most sub-radiant pole, the long-time structure is dramatic. We calculate three such frequencies [labeled in Fig.~\figurepanel{fig:time_delay_transmission_pole_10qubit_infinite}{d}] giving rise to $\text{T}=$20\%, 50\% (previously used), and 80\%, respectively, and present the result in Fig.~\ref{fig:g2_Piover2_long_time_infinite}. One can see that the long-time structure with off-resonant driving persists for more than $\Gamma t=800$ (much larger than the time of flight from one end of the array to the other without any obstacle, which is $9\pi/200\Gamma$); in contrast, the time scale of the beating is almost invisible. Moreover, as one goes from $\text{T}=20\%$ to $\text{T}=80\%$ (approaching the sub-radiant pole), this long-time scale becomes larger, as if one ``stretches'' the $g_2$ curve. In the next section we employ the concept of time delay to explain this observation.

\section{Time Delay}
\label{Sec:timedelay}

The time delay (also known as the group delay) is a way to measure the time a wavepacket spends in passing through a scattering potential \cite{deCarvalho02}.  For a symmetric potential both transmitted and reflected wavepackets are characterized by a single time delay given by 
$\uptau(k)=d\theta_k/dk$ in the general case and by 
\begin{equation}
\uptau(k)=\frac{d}{dk}\biggl(\theta_k\biggr|_{e^{ikL}\rightarrow e^{ik_0L}}\biggr)
\end{equation}
in the Markovian regime, where $\theta_k$ is the phase of the transmission amplitude $t(k)$.

A typical plot of the frequency dependence of the time delay is shown in Fig.~\ref{fig:time_delay_transmission_pole_10qubit_infinite} for $N=10$ with $k_0L=\pi/2$ and $\pi/4$. It is clear that the position and width of the peaks in the time delay are precisely captured by, respectively, the real part $\{\tilde{\omega}_i\}$ and imaginary part $\{\tilde{\Gamma}_i\}$ of the poles. This means that as one approaches the sub-radiant poles, the time delay is greatly increased. In particular, for the most sub-radiant pole we find that the time delay $\uptau$ scales as $N^3$ (fitting not shown), consistent with the finding by Tsoi \& Law that the corresponding $\tilde{\Gamma}$ scales as $N^{-3}$ \cite{TsoiPRA08}. Therefore, this feature explains the long-time structure of $g_2$ discussed in the previous section: the ``large time-scale'' is contributed by the effective decay rate of the most sub-radiant pole.

The flat structure of the time delay around $\omega_0$ can also be explained. Within the photonic band gap, single photons are mostly reflected and hence spend much less time in the qubit array. Remarkably, we find empirically that the time delay at the resonant frequency is universal,
\begin{equation}
\uptau(k=\omega_0)=\frac{2}{\Gamma} ,
\end{equation}
independent of $N$ or $L$. This is consistent with the photon simply being reflected by the first qubit encountered.

In short, the time delay is responsible for the long-time envelope of $g_2$ and it is directly connected to the simple poles of the system. In fact, for the off-resonant behavior in previous sections, our choosing to work at the frequency closest to $\omega_0$ that satisfies $\text{T}=50\%$ allowed us to take advantage of the associated long time-delay to examine nontrivial $g_2$ behavior. 

\section{Multiple Qubits in a Semi-Infinite Waveguide: Power Spectra}
\label{Sec:semi-infinite-S(w)}

We now turn to the case of a \emph{semi}-infinite waveguide and study how the presence of a mirror (the boundary) changes the response of the system. As in the infinite waveguide case above, we first focus on the power spectra (fluorescence). The $N=1$ case has been analyzed by Koshino \& Nakamura using the Heisenberg-Langevin equation (equivalent to the input-output theory) at both weak and strong coherent driving \cite{KoshinoNJP12}. We find that our L-S approach gives the same result as theirs in the weak driving limit (see Appendix~\ref{appendix:input-output theory}), which hence validates our calculation. 

Two changes in the calculation must be made for the semi-infinite case  (see Appendix~\ref{appendix:semi-infinite waveguide} for details). First, formally the Hamiltonian Eq.~\eqref{eq:Hamiltonian} remains the same, but the integration range is modified to be from negative infinity to zero. Accordingly, when solving for the single-particle eigenstate $|\phi_1(k)\rangle$, a boundary condition $t_{N}(k)+r_{N}(k)=0$ has to be imposed. We stress that in contrast to Koshino \& Nakamura's approach \cite{KoshinoNJP12}, here the boundary condition is imposed at the wavefunction level rather than the Hamiltonian level, but the results agree exactly. Secondly, as there is only one incoming and outgoing channel, the summation over the incident direction $\alpha=\{\text{R, L}\}$ must be dropped. 
As a result, adding a mirror actually reduces the number of matrix elements to be calculated. With the qubit-mirror separation defined to be $|x_N|=a$, the Markovian approximation can be employed straightforwardly by replacing $\exp(ika)$ by $\exp(ik_0a)$, as done in the infinite waveguide case.

\begin{figure}[htb]
	\centering
	\includegraphics[scale=0.95]{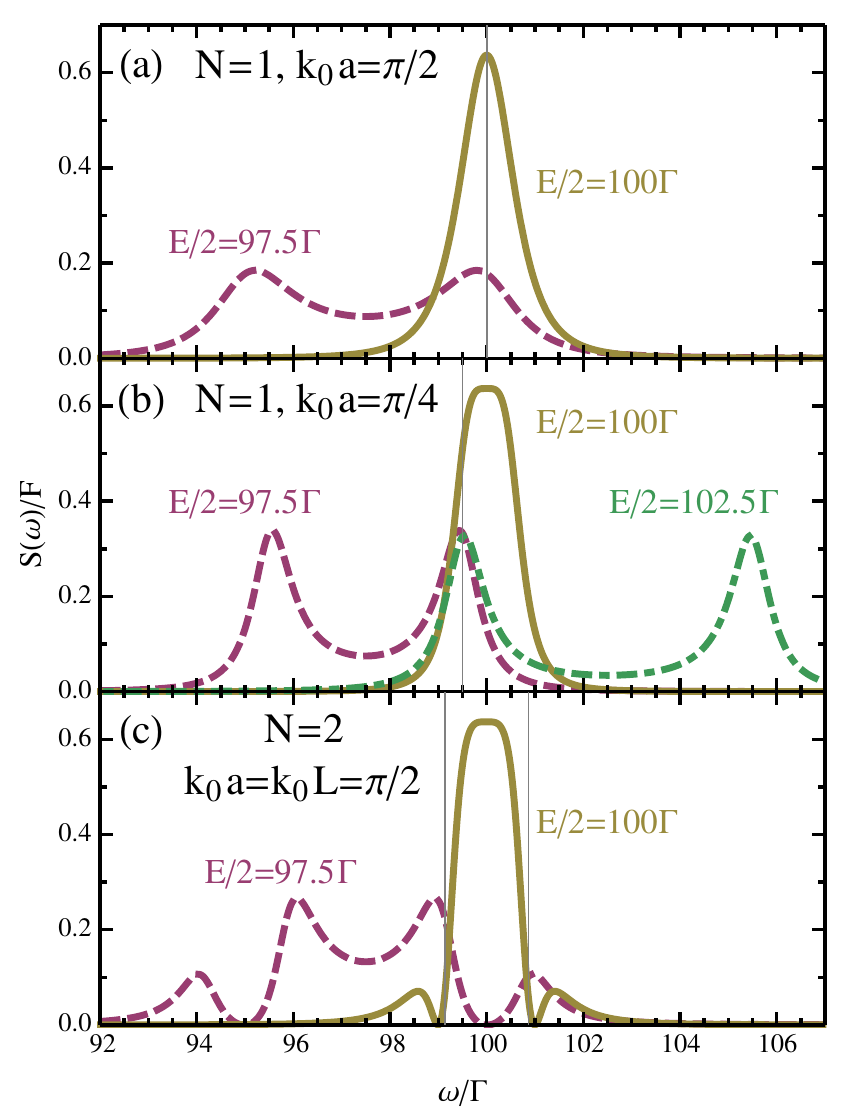}
	\caption{(Color online) Normalized power spectra of one or two qubits coupled to a semi-infinite waveguide with qubit-qubit separation $k_0L=\pi/2$ (for $N=2$). The qubit-mirror separation is $k_0a=\pi/2$ in (a) and (c), and $\pi/4$ in (b). The qubit frequency is  $\omega_0=100\Gamma$.
		\label{fig:PSD_(Piover2_Piover2)_semi_infinite}}
\end{figure}

\begin{figure}[tb]
	\centering
	\includegraphics[scale=0.9]{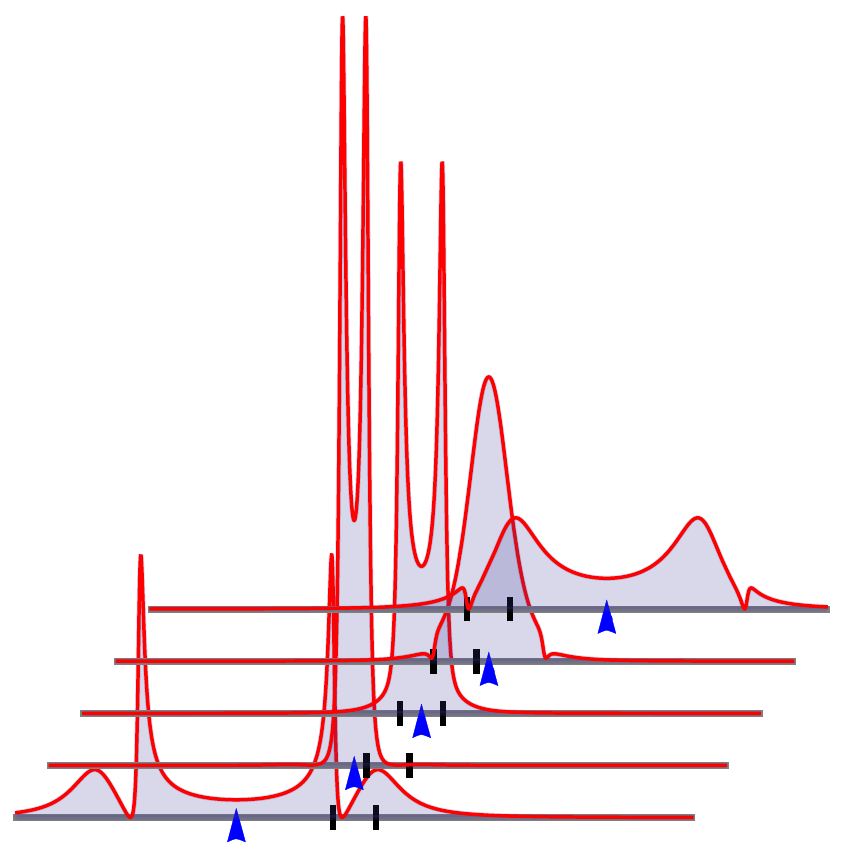}
	\caption{(Color online) Normalized power spectra of two qubits coupled to a semi-infinite waveguide with qubit-qubit separation $k_0L=\pi/2$ and qubit-mirror separation $k_0a=\pi/4$. The driving frequencies for each plot are $E/2\Gamma=96.5, 99, 100, 101, \text{and } 103.5$ (from bottom to top). The black ticks label the position of the two poles (and the qubit frequency $\omega_0=100\Gamma$ is at the center), and the blue arrow indicates the incident frequency. 
		\label{fig:PSD_(Piover2_Piover4)_semi_infinite}}
\end{figure}

In light of the discussion of the infinite waveguide case, we consider the case where the qubit-qubit separation is fixed at $k_0L=\pi/2$, allowing the distribution of poles to be symmetric for certain values of $a$, and the qubit-mirror separation is varied to see how the mirror modifies the fluorescence. We focus mostly on the case of one and two qubits, commenting on the $N=10$ results only at the end of this section. 
For one or two qubits and $k_0a=\pi/2$ or $\pi/4$, results are presented in Fig.~\ref{fig:PSD_(Piover2_Piover2)_semi_infinite} and Fig.~\ref{fig:PSD_(Piover2_Piover4)_semi_infinite}. First, note that since the reflection fluorescence \textit{is} the total fluorescence, the spectrum is always symmetric with respect to the incident frequency $E/2$. Second, since the single-photon reflection probability is always one, the way we chose the off-resonant driving frequency for the infinite case is no longer possible; instead, we have studied properties at fixed detunings.

For $N=1$ the results are similar to those of the infinite waveguide (cf. Figs.~\ref{fig:PSD_Piover2_infinite} and \ref{fig:PSD_Piover4_infinite}). Resonant driving gives a Lorentzian-like fluorescence, and off-resonant driving splits the Lorentzian peak into two. The condition for resonance is, of course, controlled by the single pole in this $N=1$ case. The main difference here compared to the infinite waveguide case is that the pole is modulated by the qubit-mirror separation $a$:
\begin{equation}
\tilde{\omega}=\omega_0-\frac{\Gamma}{2}\sin(2k_0a),\quad
\tilde{\Gamma}=\Gamma\left[1-\cos(2k_0a)\right].
\label{eq:single 2LS semi-infinite}
\end{equation}
Thus, the effective frequency and decay rate of the qubit can be changed. These relations agree with those of Ref.~\cite{KoshinoNJP12} for a hard-wall boundary condition ($\theta_b=\pi/2$ therein), and are responsible for the shift in the peak in the $k_0a=\pi/4$ case shown in Fig.~\figurepanel{fig:PSD_(Piover2_Piover2)_semi_infinite}{b}.

The spectrum changes dramatically compared with the infinite waveguide case when $N\geq2$. 
For $N=2$, the expressions for the poles are much more complicated than in the infinite waveguide case, and we do not reproduce them here. However, for the special case $k_0L=\pi/2$ we find that the poles can be simplified to
\begin{equation}
\tilde{z}_{1,2}(a)=\omega_0-\frac{i\Gamma}{2}\pm\frac{\Gamma}{2}\sqrt{1-2e^{2ik_0a}}.
\end{equation}
From this expression one can see that the ``center of mass'' is $\omega_0-i\Gamma/2$ ---it is not affected by the mirror. The two poles circulate this center in an elliptical trajectory on the complex plane as $a$ changes, in contrast to the infinite waveguide case where the two poles circulate in a perfect circle as $L$ changes \cite{TsoiPRA08, ZhengPRL13}. 

For the case $k_0a=\pi/2$, the two poles have the same decay rate and the spectra are symmetric between red and blue detuning, so only the red-detuned case is shown in Fig.~\figurepanel{fig:PSD_(Piover2_Piover2)_semi_infinite}{c}. When the driving frequency is far detuned, there are four peaks, with the inner two higher and the outer two lower, similar to that of the (total) power spectrum in the infinite case (not shown). The main difference is that here there are nodes (at which $S=0$) between the inner and the outer peaks, one of which is fixed at the bare qubit frequency $\omega_0$. As the driving frequency approaches either of the poles, the two inner peaks merge into one (a process similar to that seen in Fig.~\ref{fig:PSD_Piover2_infinite}). Next, when the driving frequency is tuned between the poles, both nodes start to be shifted and lifted, and do not touch down to zero again until the driving is on resonance ($E/2=\omega_0$).

On the other hand, for the $k_0a=\pi/4$ case the decay rates of the two poles are different, resulting in a sharper (flatter) spectrum on the red- (blue-) detuned side. To illustrate the drastically varying structure of the fluorescence, we show in Fig.~\ref{fig:PSD_(Piover2_Piover4)_semi_infinite} results for five incoming photon frequencies: substantially red-detuned, slightly red-detuned, likewise for blue-detuned, and finally on resonance. Starting from substantially red-detuned driving ($E/2=96.5\Gamma$), the four peaks and two nodes are still visible, but the right node is red-shifted away from $\omega_0$, presumably due to the asymmetric poles. As the frequency of the incoming photons is increased, the merging process happens but with one difference from the $k_0a=\pi/2$ case: the outer peaks disappear completely. For driving in between the poles, the main peak splits. In contrast to the $k_0a=\pi/2$ case, when the driving approaches the blue-detuned pole, instead of merging the two peaks actually shrink, and a single larger peak emerges between them. Finally, as the driving becomes substantially blue-detuned, the larger peak again splits into two, with the outer peaks and the nodes appearing. 

We note a special case in this progression: at $E/2=99.5\Gamma$, the entire spectrum of inelastic scattering disappears and both photons are reflected elastically. The reason is that in the steady state the wavefunctions for the two qubits differ by a phase $\pi$. Together with the phases picked up during propagation, it results in a precise destructive interference killing the photon-photon bound state.

\begin{figure}[tb]
	\centering
	\includegraphics[scale=0.9]{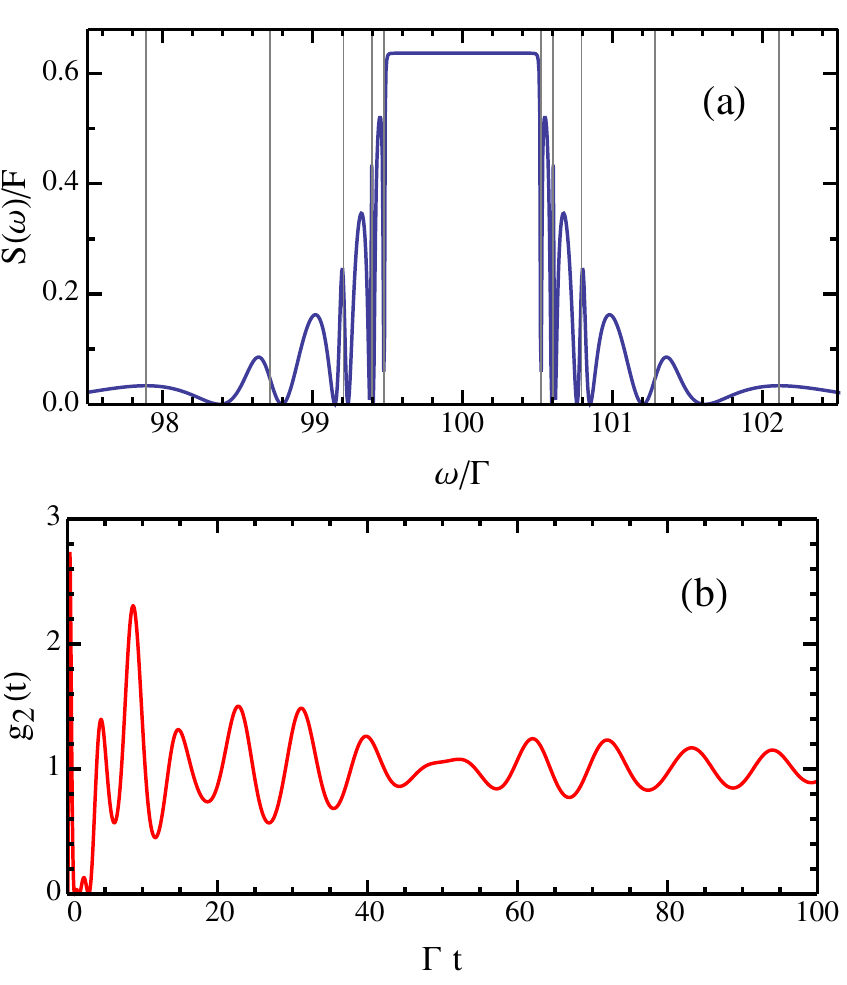}
	\caption{(Color online) A representative case of (a) $S(\omega)$ and (b) $g_2$ for 10 qubits coupled to the semi-infinite waveguide with $k_0L=k_0a=\pi/2$. The system is driven resonantly ($E/2=\omega_0=100\Gamma$). The vertical lines indicate the real parts of the poles.
		\label{fig:g2_PSD_10qubit_(Piover2_Piover2)_semi_infinite}}
\end{figure}

In Fig.~\figurepanel{fig:g2_PSD_10qubit_(Piover2_Piover2)_semi_infinite}{a} we show the power spectrum for a representative $N=10$ case. Compared to the infinite waveguide case (cf. Fig.~\ref{fig:PSD_Piover2_infinite}), note the better defined photonic band gap behavior around the 2LS resonant frequency and the sharper modulation on the sides. This comes about because the mirror effectively doubles the number of qubits that the photons see, leading to finer and stronger interference effects.

In short, adding a mirror changes drastically the spectrum of inelastic scattering by two qubits and brings in another way to modulate the distribution of the poles. More generally, this will be the case for changing the boundary condition on the semi-infinite waveguide. For superconducting qubits coupled to a microwave transmission line, while physically moving the qubit \textit{in situ} is normally not feasible, changing the boundary condition continuously with a magnetic field is readily accomplished by adding a SQUID to the end of the waveguide \cite{WilsonNat11, KoshinoNJP12}.

\section{Multiple Qubits in a Semi-Infinite Waveguide: Photon Correlations}
\label{Sec:semi-infinite-g2}

\begin{figure}[tb]
	\centering
	\includegraphics[scale=0.95]{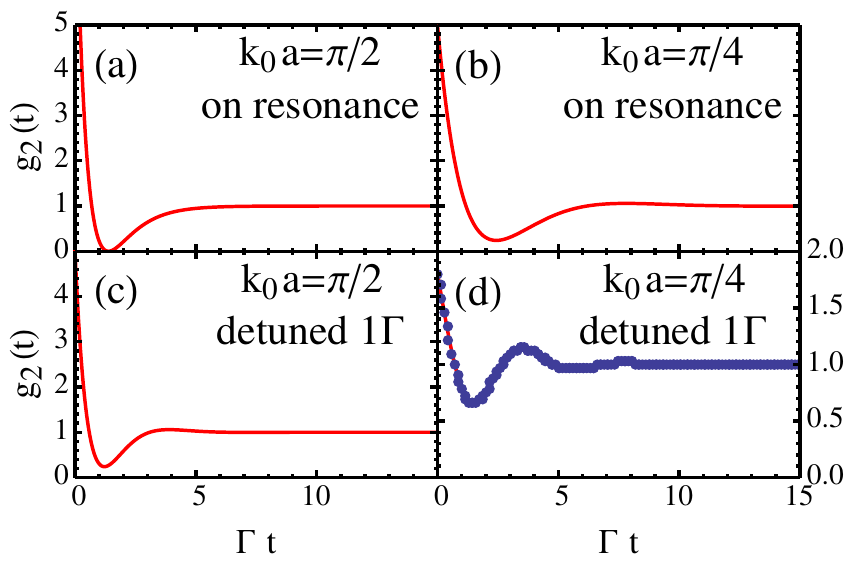}
	\caption{(Color online) $g_2$ of a single qubit coupled to a semi-infinite waveguide. The qubit-mirror separation is (a),(c) $k_0a=\pi/2$ and (b),(d) $\pi/4$. The frequency of the incoming photons is (a),(b) resonant with the 2LS ($E/2=\omega_0$) and (c),(d) detuned by $+1\Gamma$. Due to the modulated effective qubit frequency [Eq.~\eqref{eq:single 2LS semi-infinite}], for $\pi/4$ the $g_2$ with detuning $-1\Gamma$ is same as the resonant case; for $\pi/2$ the $g_2$ with detuning $-1\Gamma$ is same as the $+1\Gamma$ detuned case. The dots in panel (d) are the results of the full non-Markovian numerical calculation, and the solid curves are based on the Markovian approximation. The qubit frequency is $\omega_0=100\Gamma.$
	\label{fig:g2_Piover2_Piover4_semi_infinite}}
\end{figure}

Finally, let us turn to results for photon correlations in a semi-infinite waveguide. We first concentrate on the single-2LS case. 
Because properties are controlled by a single pole [with frequency and decay rate give in Eq.~\eqref{eq:single 2LS semi-infinite}], $g_2$ will be the same for driving frequencies equally detuned (either blue- or red-) from $\tilde{\omega}$. The result is shown in Fig.~\ref{fig:g2_Piover2_Piover4_semi_infinite} for both $k_0a=\pi/2$ and $\pi/4$ ($a=\lambda_0/4$ or $\lambda_0/8$, respectively). 

A striking difference from the infinite waveguide case is that $g_2(0)$ is no longer zero; instead, it indicates bunching in all four cases shown. This can be explained by stimulated emission: since the first photon is captured by the 2LS, the second photon passes through to the wall and is reflected back. Because of the short distance (time-of-flight $=2a/c\sim\pi/\omega_0\ll1/\Gamma$), when the second photon revisits the 2LS, the first photon has not been released, and the former can stimulate the emission of the latter, producing two photons coming out together. 

An additional difference comes from the nodes present in the wavefunciton in the semi-infinite case. We find that $\tilde{\Gamma}=0$ when $k_0a=0$, $\pi$, $2\pi$, $\cdots$, and hence no bound state is present, yielding $g_2=1$. The qubit, being placed at a node of the photonic field, is fully decoupled from the waveguide \cite{KoshinoNJP12, HoiarXiv14}.

In comparing the $k_0a=\pi/2$ results to those for $\pi/4$, it is clear that the timescale for features in $g_2$ is larger for the smaller value of $a$. That this should be the case is evident from the pole structure: they are symmetric in the $\pi/2$ case and rotated from that symmetry point for $\pi/4$. Thus the lifetime for one of the poles in the $\pi/4$ case is longer than for the $\pi/2$ poles, causing the timescale for the structure to be larger.

\begin{figure}[tb]
	\centering
	\includegraphics{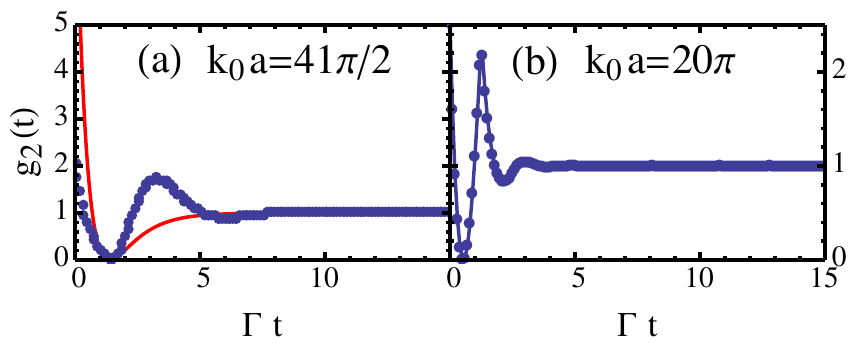}
	\caption{(Color online) $g_2$ of a single qubit coupled to a semi-infinite waveguide with resonant incoming photons. The solid (red) curves are based on the Markovian approximation while the (blue) dots result from the full non-Markovian numerical calculation. (a)~$k_0a=41\pi/2$. Note the breakdown of the Markovian approximation for this large value of $a$. (b)~$k_0a=20\pi$, thus the qubit is at a node of the single-photon wave function ($a=10\lambda_0$). In the Markovian approximation, the qubit is decoupled from the waveguide and $g_2(t)=1$. Clearly this is not the case in the full solution---there is both bunching and antibunching. The qubit frequency is $\omega_0=100\Gamma.$
		\label{fig:g2_41Piover2_20Pi_semi_infinite}}
\end{figure}

To assess the quality of the Markovian approximation, this is one of the cases we have chosen to investigate (for other results, see Fig.~\ref{fig:g2_Piover2_T50_infinite} above). In Fig.~\figurepanel{fig:g2_Piover2_Piover4_semi_infinite}{d} we compare our analytical Markovian results with the full non-Markovian numerical calculation (blue dots) when the 2LS is very near the end of the waveguide, $k_0a=\pi/4$. The agreement between the two calculations is excellent. However, as $k_0a$ becomes larger, the Markovian approximation  breaks down, as demonstrated in 
Fig.~\ref{fig:g2_41Piover2_20Pi_semi_infinite} for $k_0a=41\pi/2$ and $20\pi$ corresponding to $a=10\tfrac{1}{4}\lambda_0$ and $10\lambda_0$, respectively. This seems to happen when $a$ is larger than a few wavelengths, which for our choice of parameters means that $a$ is of order the time of flight of a photon during the decay time of the 2LS, $a\sim c/2\Gamma$. Fig.~\figurepanel{fig:g2_41Piover2_20Pi_semi_infinite}{b} shows a particularly dramatic example. If the distance between the 2LS and the mirror is small, placing the 2LS at a node of the single-particle wavefunction ($a=n\lambda_0/2$ for some $n$) causes it to be completely decoupled from the waveguide: there are no incoherently scattered photons, as has been discussed theoretically \cite{KoshinoNJP12} and seen experimentally \cite{HoiarXiv14}, and $g_2(t) = 1$ for all $t$. In contrast, for the large $a$ used in Fig.~\figurepanel{fig:g2_41Piover2_20Pi_semi_infinite}{b} so that non-Markovian effects are important, $g_2$ shows strong bunching and antibunching. Clearly, the 2LS remains coupled to the waveguide and causes nonlinear bound state effects. Though the parameters considered in Fig.~\ref{fig:g2_41Piover2_20Pi_semi_infinite} fit the discussion of non-Markovianity in Ref.~\cite{TufarelliPRA14} in terms of the qubit excitation, we leave the problem of a quantitative characterization of the non-Markovianity in this system for further study.



\begin{figure}[tb]
	\centering
	\includegraphics[scale=0.95]{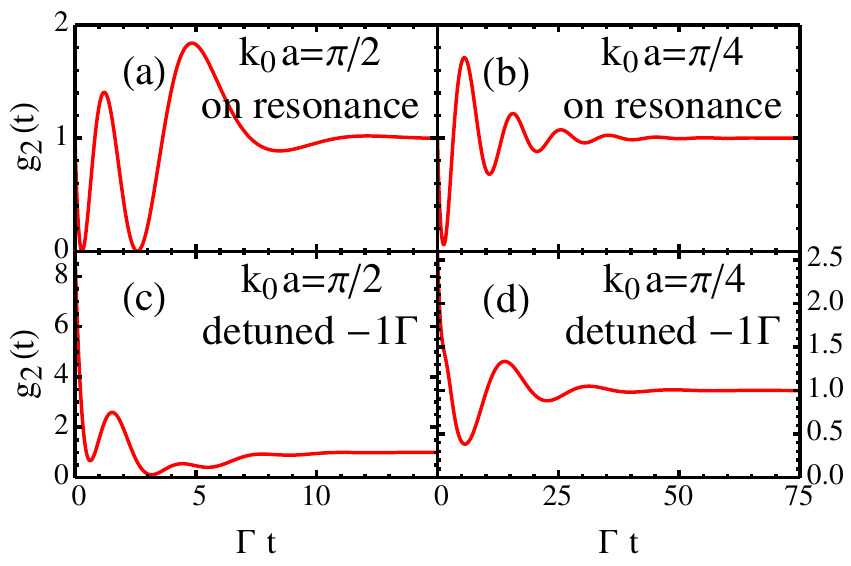}
	\caption{(Color online) $g_2$ of two qubits coupled to a semi-infinite waveguide with qubit-qubit separation $k_0L=\pi/2$. The qubit-mirror separation in the first column is $k_0a=\pi/2$ and in the second $\pi/4$. The first row has resonant driving ($E/2=\omega_0=100\Gamma$) and the second is detuned by $-1\Gamma$.
		\label{fig:g2_2qubit_(Piover2_Piover2)_(Piover4_Piover2)_semi_infinite}}
\end{figure}

Results for a two-qubit case are shown in Fig.~\ref{fig:g2_2qubit_(Piover2_Piover2)_(Piover4_Piover2)_semi_infinite}. We again use $k_0L=\pi/2$, giving rise to a symmetric pole distribution in an infinite waveguide, and focus on the effect of the mirror. As expected, the oscillation when $k_0a=\pi/4$ lasts longer than that with $\pi/2$ due to the existence of the sub-radiant pole. This result is consistent with the finding from the calculation of power spectra (Figs.~\ref{fig:PSD_(Piover2_Piover2)_semi_infinite} and \ref{fig:PSD_(Piover2_Piover4)_semi_infinite}). As $N$ increases, we find that the behavior of $g_2$ can be explained in much the same way as in the infinite waveguide situation by examining the pole distribution. 

We show one representative example of $g_2$ for 10 qubits and resonant driving in  Fig.~\figurepanel{fig:g2_PSD_10qubit_(Piover2_Piover2)_semi_infinite}{b}. While qualitatively similar to the result for an infinite waveguide (cf. Fig.~\ref{fig:g2_Piover2_Piover4_on_resonance_infinite}), $g_2$ here shows a more complex interference pattern and stronger modulation, as for the power spectrum.  

\section{Conclusion}

In this work we have surveyed a wide variety of multi-qubit waveguide-QED structures, focusing on their two-photon nonlinearities as manifested in the power spectrum [resonance fluorescence, $S(\omega)$] and photon correlation function [second-order coherence, $g_2(t)$]. It is clear that in the multi-qubit case (we studied from one to ten qubits), these two functions show a great deal of structure caused by the interference of the partial waves scattering from different combinations of qubits. Given that oscillations are ubiquitous in $g_2(t)$ here, the initial correlation $g_2(t=0)$ certainly cannot be used as an indication of whether the system generally causes bunching or anti-bunching of photons.

The structure in $g_2(t)$ and $S(\omega)$ generally becomes sharper as the number qubits, $N$, increases---an effect particularly noticeable in the resonance fluorescence, see Figs.~\ref{fig:PSD_Piover2_infinite} and \ref{fig:PSD_Piover4_infinite}. This is natural as the interference effects become more complicated and the photonic band gap builds up. In $g_2(t)$ the deviations from semi-classics ($g_2=1$) persist for a much longer time than one might initially expect, and this time increases upon increasing the number of qubits. For $N=10$ the decay of correlation in time is very slow indeed (see Fig.~\ref{fig:g2_Piover2_T50_infinite}).

Many of the features and trends in our results can be roughly explained by referring to the poles of the transmission amplitude. These poles  (see Fig.~\ref{fig:time_delay_transmission_pole_10qubit_infinite} for an example) also appear in the single particle Green function used in calculating the correlation or ``bound state'' effects.
We have seen that the most sub-radiant pole is especially important. The ubiquitous oscillations seen come from beating between the frequency of the incoming photons (driving frequency) and the real part of the most sub-radiant pole. Other oscillations no doubt come from beating among the different poles and between them and the driving frequency. The long decay time, seen especially for large $N$, comes from the small decay rate of the most sub-radiant pole; we saw that this scale also appears as the time delay.

Some notable features in our results include: The total power spectrum is symmetric about the driving frequency, but note that the spectrum of only the transmitted or reflected photons (in the infinite waveguide case) are not. We have seen that there is often either bunching or anti-bunching in \emph{both} transmission and reflection---because the photons can spend a significant amount of time traveling among the different qubits, it is not the case that if one is bunched the other should be anti-bunched.
It is unfortunate that there are very few trends as the number of qubits increases. One exception is the interesting case in which there is strong anti-bunching in transmission and bunching in reflection that lasts for a long time (Fig.~\ref{fig:g2_Piover4_T50_infinite}); this is enhanced as $N$ increases due to the increasingly sub-radiant pole produced by the multiple interference.

The infinite and semi-infinite waveguide cases show a number of differences. Perhaps the most important is that a single 2LS can cause bunching of two photons in the semi-infinite case (Fig.~\ref{fig:g2_Piover2_Piover4_semi_infinite}) while in the infinite waveguide case there must be complete anti-bunching [$g_2(0)\!=0$]. The mirror in the semi-infinite case acts to effectively double the number of qubits, and so there is more sharp structure in the presence of a mirror for the same number of qubits.

The effects of loss and dephasing have been entirely neglected in the present study; what effect would they have? Let $\Gamma'$ denote the rate of decay of one of the qubits to modes other than the waveguide. Then, one expects that any structure on a timescale larger than $(\Gamma')^{-1}$ will be smoothed out. In particular, phenomena related to the most sub-radiant pole will disappear first, when $\Gamma' > \tilde{\Gamma}$. Pure dephasing causes a similar smoothing of interference effects without, of course, relaxing the excited state population. In addition to smoothing, dephasing can cause the power spectrum to be asymmetric about the input frequency \cite{KoshinoNJP12}. One may think, then, that most of the structure in our calculated curves would disappear. However, there has been tremendous experimental progress recently toward making systems whose loss rate is very low and whose dephasing is even smaller. Purcell factors, defined by $\Gamma/\Gamma'$, greater than ten have been demonstrated in more than one experimental platform: for superconducting qubits coupled to a microwave transmission line \cite{HoiNJP13, KoshinoPRL13, vanLooScience13, HoiarXiv14}, for instance, as well as quantum dots coupled to plasmonic nanostructures \cite{AkselrodNatPho14} and to a photonic waveguide \cite{LauchtPRX12, ArcariPRL14}. Given the rapid pace at which the experimental systems are advancing, we think considering a system in which $P \sim 100$ is reasonable; for such a Purcell factor, the large majority of the effects presented here will survive.

We have considered the effect of weak disorder on our results, using $\pm 5\%$ as a typical experimental variation in, for instance, the qubit frequency among nominally identically made qubits. Such a weak disorder does little to change the results. Increasing the disorder would, of course, change the interference effects such that the average or typical results would show much less structure. However, a large variation in the qubit and coupling properties seems unreasonable for the experimental systems studied, and in addition, because the system is 1D, localization of the wavefunctions appears immediately, producing very sharp resonances.

The large majority of our results were obtained in the Markovian approximation, as this is the case relevant to most current experiments. We have compared to a full non-Markovian calculation in a few cases [for example, see Figs.~\figurepanel{fig:g2_Piover2_T50_infinite}{d}, \figurepanel{fig:g2_Piover2_Piover4_semi_infinite}{d}, and \ref{fig:g2_41Piover2_20Pi_semi_infinite}]. If the spacing of the qubits is small, then the agreement between the two is very good. However, for large separation, we see a substantial difference between the two results. For instance, for one qubit coupling to a semi-infinite waveguide, placing the qubit at a node of the single particle wavefunction does not lead to a decoupling of the qubit (as it would in the Markovian regime). The criterion for when these effects set in is that the separation $L$ should be larger than the distance a photon can travel in the decay time of a qubit: if $L > c/\Gamma$, a non-Markovian calculation is required. 
Experimentally the non-Markovian regime may be reached by connecting superconducting circuits using a coaxial cable \cite{RochPRL14}. 

We end by simply noting the richness of the correlation phenomena in these waveguide-QED systems and the bright future for further investigations because of rapid experimental progress. The methods that we use here to study small ensembles of two-level systems could be extended in a straightforward way to more complex systems, such as three- or $N$-level systems and photons of different frequency.

\acknowledgements
We thank H.~Zheng, D.~Gauthier, I.-C. Hoi, K.~Lalumi\`{e}re, A.~Blais and F.~Ciccarello for valuable disucssions. We also thank the computing support and resources provided by Duke Research Computing and the Open Science Grid \cite{OSGreference1, OSGreference2}; the latter is supported by the National Science Foundation and the U.S. Department of Energy's Office of Science. Most figures presented in this paper were prepared using the SciDraw scientific figure preparation system \cite{*[{ \href{http://scidraw.nd.edu}{SciDraw} is the successor of the LevelScheme package; for futher detail, see }]  [{.}] CaprioCompPhyComm05}. This work was supported by U.S. NSF Grant No.~PHY-14-04125.

\appendix
\section{Matrix elements}
\label{appendix:matrix elements}

In this appendix we follow the procedure and notation of our previous work  Refs.~\cite{ZhengPRL13, FangEPJQT14} to calculate the matrix elements needed for the (inelastic) power spectra. We focus on the calculation for the infinite waveguide cases, and leave the semi-infinite cases for the next appendix. Readers interested in details of constructing the two-photon wavefunction $|\psi_2\rangle$ should refer to these references. 

As discussed in the main text, the goal is to compute the matrix element $_{\alpha'}\langle\phi_1(k)|a_\alpha(x)|\psi_2\rangle$.  Starting from the single-particle eigenstate 
\begin{align}
|\phi_1(k)\rangle_\alpha= & \bigg[\int dx \left(\phi_\text{R}^\alpha(k,x) a_\text{R}^\dagger(x)+\phi_\text{L}^\alpha(k,x) a_\text{L}^\dagger(x)\right)
\nonumber \\
& \quad +\sum_{i=1}^N e_i^\alpha(k)d_i^\dagger\bigg]|0\rangle
\label{eq:single particle eigenstate}
\end{align}
with the incident photon of momentum $k$ propagating in the $\alpha$-direction, one can construct the two-photon direct-product plane-wave state 
\begin{equation}
|\phi_2(k_1, k_2)\rangle_{\alpha_1, \alpha_2}=\frac{1}{\sqrt{2}}
\biggl(|\phi_1(k_1)\rangle_{\alpha_1} \otimes |\phi_1(k_2)\rangle_{\alpha_2}\biggr)
\end{equation}
and the associated two-particle identity operator $\mathcal{I}_2=\sum_{\alpha_1,\alpha_2}\int dk_1dk_2 |\phi_2(k_1, k_2)\rangle_{\alpha_1,\alpha_2}\langle \phi_2(k_1,k_2)|$.
By inserting $\mathcal{I}_2$ into Eq.~\eqref{eq:two-photon wavefunction}, one observes that the entire two-photon wavefunction $|\psi_2\rangle$ can be expressed using $|\phi_2\rangle$, so the problem is reduced to calculating the matrix element $_{\alpha'}\langle\phi_1(k)|a_\alpha(x)|\phi_2\rangle$; the double-momentum integral mentioned in the main text is introduced during this insertion of $\mathcal{I}_2$. 

By employing the definition of $|\phi_1\rangle$, one observes that
\begin{equation}
a_\text{R}(x)|\phi_2(k_1, k_2)\rangle_{\alpha_1,\alpha_2}=\biggl(\frac{\phi_\text{R}^{\alpha_1}(k_1,x)}{\sqrt{2}}|\phi_1(k_2)\rangle_{\alpha_2}
+(2\leftrightarrow1)\biggr),
\label{a phi_2}
\end{equation}
and hence 
\begin{widetext}
	\begin{subequations}
		\label{eq:plane wave matrix element}
		\begin{align}
		_\text{R}\langle \phi_1(k)|a_\text{R}(x)|\phi_2(k_1, k_2)\rangle_{\text{R,R}}&=\frac{1}{\sqrt{2}}\biggl(\phi_\text{R}^\text{R}(k_1,x)\delta(k-k_2)+\phi_\text{R}^\text{R}(k_2,x)\delta(k-k_1)\biggr) \\
		_\text{R}\langle \phi_1(k)|a_\text{R}(x)|\phi_2(k_1, k_2)\rangle_{\text{L,R}}&=\frac{1}{\sqrt{2}}\phi_\text{R}^\text{L}(k_1,x)\delta(k-k_2) \\
		_\text{R}\langle \phi_1(k)|a_\text{R}(x)|\phi_2(k_1, k_2)\rangle_{\text{R,L}}&=\frac{1}{\sqrt{2}}\phi_\text{R}^\text{L}(k_2,x)\delta(k-k_1) \\
		_\text{R}\langle \phi_1(k)|a_\text{R}(x)|\phi_2(k_1, k_2)\rangle_{\text{L,L}}&=0\\
		_\text{L}\langle \phi_1(k)|a_\text{R}(x)|\phi_2(k_1, k_2)\rangle_{\text{R,R}}&=0 \\
		_\text{L}\langle \phi_1(k)|a_\text{R}(x)|\phi_2(k_1, k_2)\rangle_{\text{L,R}}&=\frac{1}{\sqrt{2}}\phi_\text{R}^\text{R}(k_2,x)\delta(k-k_1) \\
		_\text{L}\langle \phi_1(k)|a_\text{R}(x)|\phi_2(k_1, k_2)\rangle_{\text{R,L}}&=\frac{1}{\sqrt{2}}\phi_\text{R}^\text{R}(k_1,x)\delta(k-k_2) \\
		_\text{L}\langle \phi_1(k)|a_\text{R}(x)|\phi_2(k_1, k_2)\rangle_{\text{L,L}}&=\frac{1}{\sqrt{2}}\biggl(\phi_\text{R}^\text{L}(k_1,x)\delta(k-k_2)+\phi_\text{R}^\text{L}(k_2,x)\delta(k-k_1)\biggr).
		\end{align}
	\end{subequations}
	One can compute $_{\alpha}\langle\phi_1(k)|a_\text{L}(x)|\phi_2\rangle_{\alpha_1, \alpha_2}$ in a similar way.
	Before giving the final result, we find that defining the following four functions ($RR_i, RL_i, LR_i, LL_i$) is useful:
	\begin{equation}
	\alpha\beta_i(k, x) \equiv\sum_{\alpha_1, \alpha_2}\int dk_1'dk_2'
	\frac{_{\alpha}\langle\phi_1(k)|a_\beta(x)|\phi_2(k_1', k_2')\rangle_{\alpha_1, \alpha_2}\langle\phi_2(k_1', k_2')|d_id_i\rangle}{E-(k_1'+k_2')+i\varepsilon},\quad
	\alpha, \beta = R, L.
	\end{equation}
\end{widetext}

Collecting all the pieces together, the target matrix element is given by
\begin{align}
&_\alpha\langle\phi_1(k) | a_\beta(x) | \psi_2(k_1, k_2)\rangle_\text{RR}=\!
_\alpha\langle\phi_1(k) | a_\beta(x) | \phi_2(k_1, k_2)\rangle_\text{RR}\nonumber\\
& \quad - \sum_{i,j=1}^N \alpha\beta_i(k, x)\left(G^{-1}\right)_{i,j}
\langle d_jd_j | \phi_2(k_1, k_2)\rangle_\text{RR}.\label{eq:two-photon matrix element}
\end{align}
From Eq.~\eqref{eq:plane wave matrix element} and \eqref{eq:two-photon matrix element}, it is clear that any term multiplying the first term above will carry a Dirac delta function, and so it will not contribute to the inelastic power spectrum. Only the product appearing in the second term can contribute. In evaluating that contribution, we find that the following triple integrals ($T^{RR}, T^{LR}, T^{RL}, T^{LL}$) simplify the final result:
\begin{align}
T^{\alpha R}_{i, j}(E, \omega)&=\!\int_0^\infty\!\! dt\, e^{-i\omega t}\int \!dk\,
\alpha R_i^*(k,  x_0)  \alpha R_j(k,  x_0+t),\nonumber\\
T^{\alpha L}_{i, j}(E, \omega)&=\!\int_{-\infty}^0\!\! dt\, e^{i\omega t}\int \!dk\,
\alpha L_i^*(k,  x_0)  \alpha L_j(k,  x_0-t)
\end{align}
for $\alpha=R, L$. We note two things here: First, we call the objects $T^{\alpha\beta}_{i, j}$ triple integrals because there are three momentum integrals to be done; the time integration can be evaluated trivially. Second, although formally the function $\alpha\beta_i$ carries an $ x_0$ dependence, requiring the position $ x_0$ of the detector to be away from the qubit array ($ x_0\gg0$ for transmission and $ x_0\ll0$ for reflection), as discussed in the main text, will remove this dependence, as expected.

Finally, substituting Eq.~\eqref{eq:two-photon matrix element} into Eq.~\eqref{eq:power_spectrum_in_practice} and removing all terms proportional to delta functions after integration (they represent coherent scattering as discussed in the main text), we obtain the incoherent power spectrum 
\begin{widetext}
	\begin{equation}
	S_\beta(\omega)=2\sum_{i, j, k, l}\re\left[ _{\textrm{RR}}\langle \phi_2(k_1, k_2)|d_id_i\rangle
	(G^{-1})_{i,j}^*\left(T^{R\beta}+T^{L\beta}\right)_{j,k}(G^{-1})_{k,l}
	\langle d_l d_l|\phi_2(k_1, k_2)\rangle_{\textrm{RR}} \right],
	\end{equation} 
	where $\beta=R$ is for the transmission fluorescence and $\beta=L$ for the reflection fluorescence. In the derivation we use the property that the transpose of the matrix $G$ is itself.
	
	To wrap up, we emphasize two things: (a) $k_1=k_2=E/2$ is used, and (b) the Markovian approximation can be introduced at the stage of the evaluation of the matrices $G$ and $T$. For a single-qubit coupled to an infinite waveguide, the double and triple integrations can be done exactly, and the result is 
	\begin{equation}
	S_R(\omega)=S_L(\omega)=\frac{\Gamma^4}{4\pi^2}\frac{1}{\left[(E-\omega_0-\omega)^2+\Gamma^2/4\right]\left[(E/2-\omega_0)^2+\Gamma^2/4\right]\left[(\omega-\omega_0)^2+\Gamma^2/4\right]},\label{eq:single qubit S(w) infinite waveguide}
	\end{equation}
	which gives the $N=1$ plot in Fig.~\ref{fig:PSD_Piover2_infinite}. For $N\geq2$ the results are more complicated, and we do not reproduce them here. In particular, for $N\geq5$ it is well-known that there is no explicit formula for the roots of a degree $N$ polynomial, so we use Mathematica to symbolically keep track of all the poles and to calculate the measurables presented in this paper.

	\section{Modifications for the Semi-Infinite Waveguide}
	\label{appendix:semi-infinite waveguide}
	
	As mentioned in the main text the calculation of the semi-infinite waveguide cases is actually simpler because all the incident light can only propagate to the right, be reflected, and then collected at one end. As the first consequence, in contrast to Eq.~\eqref{eq:single particle eigenstate}, the single-particle eigenstate $|\phi_1\rangle$ no longer carries a directional index $\alpha$:
	\begin{equation}
	|\phi_1(k)\rangle=\left[\int_{-\infty}^0 dx \left(\phi_\text{R}(k,x) a_\text{R}^\dagger(x)+\phi_\text{L}(k,x) a_\text{L}^\dagger(x)\right)+\sum_{i=1}^N e_i(k)d_i^\dagger\right]|0\rangle.
	\end{equation}
	Similar to the infinite waveguide cases, to solve for the photon and qubit wavefunctions, $\phi_\text{R}(k,x), \phi_\text{L}(k,x)$, and $\{e_i(k)\}$, the following ansatz is used:
	\begin{subequations}
		\label{eq:single particle eigenstate semi-infinite}
			\begin{align}
			\phi_\text{R}(k,x) &= \frac{e^{ikx}}{\sqrt{2\pi}}\left(\theta(x_1-x)+\sum_{i=1}^{N-1} t_i(k)\theta(x-x_i)\theta(x_{i+1}-x) 
			+ t_N(k)\theta(x-x_N)\right)\!,\\
			\phi_\text{L}(k,x) &= \frac{e^{-ikx}}{\sqrt{2\pi}}\left(r(k)\theta(x_1-x)+\sum_{i=1}^{N-1} r_i(k)\theta(x-x_{i})\theta(x_{i+1}-x)+ r_N(k)\theta(x-x_N)\right),
			\end{align}
	\end{subequations}
\end{widetext}
where $r(k)$ is the single-photon reflection amplitude and $|r(k)|^2=1$ could serve as a check of the calculation. Since the number of equations given by the Schrodinger equation does not match the number of unknowns, we need one more equation to close the set, which is the boundary condition at $x=0$. In this work, we use hard-wall boundary conditions,
\begin{equation}
\phi_\text{R}(k,0)+\phi_\text{L}(k,0)=0 \quad \Rightarrow \quad t_N(k)+r_N(k)=0,
\end{equation}
where $x=0$ is the mirror position. Following the standard procedure (see e.g.\ Ref.~\cite{TsoiPRA08}) one is then able to compute the wavefunctions. 

For the two-photon wavefunction the generalization is straightforward. The only change, also stated in the main text, is that the single- and double-particle identity operators, $\mathcal{I}_1$ and $\mathcal{I}_2$, carry no directional indexes. In contrast to Eq.~\eqref{eq:plane wave matrix element}, in this case the only matrix element needed is $\langle \phi_1(k)|a_\text{L}(x)|\phi_2(k_1, k_2)\rangle$. Therefore, actually one only needs to calculate the $T^{RL}$ matrix for the power spectra. This again shows  the simplicity of the semi-infinite waveguide cases. 

For a single qubit coupled to a semi-infinite waveguide, unlike the infinite case, however, the Markovian approximation is necessary if one wants to analytically evaluate the integrals, and the result is the same as Eq.~\eqref{eq:single qubit S(w) infinite waveguide} (up to an irrelevant prefactor gone after normalization), except that the qubit frequency and decay rate are replaced by Eq.~\eqref{eq:single 2LS semi-infinite}. This result agrees with Ref.~\cite{KoshinoNJP12} in the weak driving limit, as we discuss next in Appendix \ref{appendix:input-output theory}.

\section{Heisenberg-Langevin Equations: \\ One \& Two Qubits}
\label{appendix:input-output theory}

The Heisenberg-Langevin (H-L) equation \cite{GardinerQN00}, used in Ref.~\cite{KoshinoNJP12}, is useful for studying a system under arbitrarily large coherent driving. The Langevin (``noise'') term enters the equation by integrating out the field degree of freedom and carries the information of the driving strength via the initial state. In this sense, the H-L equation is formally equivalent to input-output theory and gives the same set of differential equations \cite{GardinerQN00}. 

Specifically, to calculate the measurables using the H-L equation, the strategy is to write down the equations of motion for both atomic and photonic operators, and then to integrate out the photonic degree of freedom, leaving a set of first-order differential equations for qubits only. Then one solves the dynamics of the qubits so obtains the steady state of the system. Substituting the solution back into the formal solution of the field gives the photon dynamics. The calculation is explained in detail in Ref.~\cite{KoshinoNJP12}, so we just list the key changes to arrive at the results. Here we focus on one- and two-qubits coupled to an infinite waveguide as examples; the generalization to multiple qubits is straightforward \cite{LalumierePRA13, CanevaarXiv15}
but cumbersome.

The first step is to get the formal solutions of the photonic operators in real space from the Hamiltonian Eq.~\eqref{eq:Hamiltonian}:
\begin{widetext}
	\begin{align}
	a_\text{R}(x,t)&=a_\text{R}(x-t,0)-iV\sum_{i=1}^N\sigma_{i-}(t-x+x_i)\Theta(x_i<x<t+x_i),\nonumber\\
	a_\text{L}(x,t)&=a_\text{L}(x+t,0)-iV\sum_{i=1}^N\sigma_{i-}(t+x-x_i)\Theta(-t+x_i<x<x_i),\label{eq:HL-FormalSolution}
	\end{align}
\end{widetext}
where the generalized step function is defined by
\begin{equation}
\Theta(a<x<b)=\begin{cases}
1,&\quad a<x<b \\
1/2,&\quad x=a \text{ or } x=b\\
0,&\quad\text{otherwise},
\end{cases}
\end{equation}
and we have chosen the initial time to be $t=0$. Note that during the derivation it is safe to define the Fourier transform of $a_\text{R/L}(x)$ because of the RWA.

\subsection{One qubit}
\label{appendix:input-output theory one-qubit}

Next, one substitutes these formal solutions into the qubits' equations of motion. For a single qubit the calculation is straightforward, and the result is given by (cf. Ref.~\cite{KoshinoNJP12})
\begin{widetext}
	\begin{equation}
	\frac{d}{dt}\begin{pmatrix} s_1 \\ s_1^* \\ s_2 \end{pmatrix}=
	\begin{pmatrix}
	-\Gamma/2+i\delta\omega & 0 & i\Omega \\
	0 & -\Gamma/2-i\delta\omega & -i\Omega \\
	i\Omega/2 & -i\Omega/2 & -\Gamma
	\end{pmatrix}
	\cdot\begin{pmatrix} s_1 \\ s_1^* \\ s_2 \end{pmatrix}
	+\begin{pmatrix} -i\Omega/2 \\ i\Omega/2 \\ 0 \end{pmatrix},\label{eq:1qubitHLequation}
	\end{equation}
	where $s_1(t)=\langle\sigma_-(t)\rangle \exp(ikt)$, $s_2(t)=\langle \sigma_+(t)\sigma_-(t)\rangle$, $\delta\omega=k-\omega_0$, $\langle O(t)\rangle$ represents the expectation value of the operator $O$ at time $t$, and $\Omega=\sqrt{2\Gamma}A$ (note that we use $k=E/2$ and $A$ to represent the driving frequency $\omega_p$ and driving strength $E$ of Ref.~\cite{KoshinoNJP12}). We emphasize that the equations above can be cast into a compact form: $\partial_t \mathbf{S}=D\mathbf{S}+\mathbf{F}$ with matrix $D$ and vector $\mathbf{F}$ being constant in time. This $D$-matrix is of great importance because it allows one to calculate all higher-order correlation functions using the quantum regression theorem (making the Markovian approximation) \cite{GardinerQN00}. 
	
	
	Assuming that the qubit is initially in the ground state, one finds $s_1(0)=s_2(0)=0$, and so the steady state given by Eq.~\eqref{eq:1qubitHLequation} is
	\begin{align}
	s_1(\infty)&=-\frac{i}{2}\frac{\Gamma(\Gamma/2+i\delta\omega)\Omega}{\Gamma(\Gamma^2/4+\delta\omega^2)+\Gamma\Omega^2/2},\nonumber\\
	s_2(\infty)&=\frac{1}{2}\frac{\Gamma\Omega^2}{\Gamma(\Gamma^2/4+\delta\omega^2)+\Gamma\Omega^2/2}.\label{eq:1qubitSteadyState}
	\end{align}
	Comparing this solution to Eqs.~(30) and (31) of Ref.~\cite{KoshinoNJP12}, one can see that in general they possess the same form, but there the decay rates and the qubit frequency are ``renormalized'' to the values in Eq.~\eqref{eq:single 2LS semi-infinite} because of the mirror. 
	
	Given the qubit steady states, one can compute the transmission and reflection amplitudes for weak coherent driving. For the transmitted part, we take the observation time $T\gg  x_0 >0$ and use Eq.~\eqref{eq:HL-FormalSolution} to look at the right-going photons:
	\begin{align}
	\langle a_\text{R}( x_0, T)\rangle &=\langle a_\text{R}( x_0-T, 0)\rangle-iV\langle\sigma_-(T- x_0)\rangle\Theta(0< x_0<T)\nonumber\\
	&=\left(1-\frac{1}{2}\sqrt{\frac{\Gamma}{2}}\frac{\Gamma(\Gamma/2+i\delta\omega)2\sqrt{\Gamma/2}}{\Gamma(\Gamma^2/4+\delta\omega^2)+\Gamma^2 A^2}\right)A e^{ik( x_0-T)}.
	\end{align}
\end{widetext}
The transmission amplitude for weak coherent driving can be defined as
\begin{equation}
t(k)\equiv\frac{\langle a_\text{R}( x_0, T)\rangle}{A e^{ik( x_0-T)}}\xrightarrow{A\rightarrow0}\frac{k-\omega_0}{k-\omega_0+i\Gamma/2}.
\label{eq:tk_def}
\end{equation}
Similarly, the reflection amplitude for weak coherent driving is ($ x_0<0, T\gg| x_0|$)
\begin{equation}
r(k)\equiv\frac{\langle a_\text{L}( x_0, T)\rangle}{A e^{-ik( x_0+T)}}\xrightarrow{A\rightarrow0}\frac{-i\Gamma/2}{k-\omega_0+i\Gamma/2},
\label{eq:rk_def}
\end{equation}
and the qubit steady state is
\begin{equation}
e(k)\equiv\frac{s_1(\infty)}{A}\xrightarrow{A\rightarrow0}\frac{\sqrt{2\pi}\sqrt{\Gamma/4\pi}}{k-\omega_0+i\Gamma/2}. 
\label{eq:ek_def}
\end{equation}
These results agree with previous studies 
\cite{ShenPRL05, ChangNatPhy07, ZhengPRA10} (the factor $\sqrt{2\pi}$ in $e(k)$ comes from the different definition of input state for two formalisms). We emphasize that the definitions 
(\ref{eq:tk_def})-(\ref{eq:ek_def})
are valid only for weak coherent driving; for arbitrary driving amplitude $A$, in general $|\langle a_\text{R}\rangle|^2+|\langle a_\text{L}\rangle|^2\neq A^2$.

Let us now turn to calculating the power spectrum. While for a single qubit this calculation is straightforward, we sketch how to employ the quantum regression theorem to simplify the calculation for more complicated cases. Suppose we define 
$s_3(t')=\langle\sigma_+(T)\sigma_-(T+t')\rangle e^{ik t'}$, 
$s_4(t')=\langle\sigma_+(T)\sigma_+(T+t')\rangle e^{-ik(2T+ t')}$, and
$s_5(t')=\langle\sigma_+(T)\sigma_+(T+t')\sigma_-(T+t')\rangle e^{-ik T}$,
then the quantum regression theorem states that the vectors 
\begin{equation}
\mathbf{S}(t')=\begin{pmatrix} s_3 \\ s_4 \\ s_5 \end{pmatrix},\quad
\mathbf{F}=\begin{pmatrix} -i\Omega s_1^*(\infty)/2 \\ i\Omega s_1^*(\infty)/2 \\ 0 \end{pmatrix}
\end{equation}
also satisfy $\partial_{t'} \mathbf{S}=D\mathbf{S}+\mathbf{F}$. As a result, the matrix $D$ is a sort of the characteristic of the system: it gives a closed set of first-order differential equations by properly defining the vectors $\mathbf{S}$ and $\mathbf{F}$.

Furthermore, in this compact notation we can separate the vector $\mathbf{S}(t)$ into two parts: $\mathbf{S}(t)=\mathbf{\boldsymbol\delta S}(t)+\mathbf{S}(\infty)$, where the former represents the transit dynamics that decays to zero and the latter represents the steady state. One can immediately see that the former instead satisfies a homogeneous first-order differential equation: $\partial_t \mathbf{\boldsymbol\delta S}=D\mathbf{\boldsymbol\delta S}$. If we define a new vector $\mathbf{I}$ by
\begin{equation}
\mathbf{I}(\omega)=\int_0^\infty dt'\, e^{i(\omega-k)t'}\mathbf{\boldsymbol\delta S}(t'),
\end{equation} 
then, after integration by parts, the vector $\mathbf{I}$ is given by
\begin{equation}
\mathbf{I}(\omega)=-\left(D+i(\omega-k)\mathcal{I}\right)^{-1}\mathbf{\boldsymbol\delta S}(0),
\end{equation}
where $\mathcal{I}$ is the identity matrix and $\mathbf{\boldsymbol\delta S}(0)=\mathbf{S}(0)-\mathbf{S}(\infty)$. This shows that by computing the vector $\mathbf{I}$ one can obtain the Fourier-transform of the two-time correlations. 

After some algebra we arrive at the transmission power spectrum
\begin{align}
S_\text{R}(\omega)&=\left[A^2+2A\sqrt{\frac{\Gamma}{2}}\re\left(i s_1^*(\infty)\right)+\frac{\Gamma}{2}|s_1(\infty)|^2\right]\delta(\omega-k)\nonumber\\
&\quad+\frac{\Gamma}{2\pi} \re I_3(\omega),
\end{align}
where the first term proportional to the delta function represents the coherent scattering, and the second term represents the incoherent scattering.  
Note that knowing $I_3(\omega)$ (the component of the vector $\mathbf{I}$ corresponding to $s_3$) is enough to determine the incoherent power spectra [see Ref.~\cite{KoshinoNJP12} for an explicit expression for $I_3(\omega)$].
The reflection power spectrum is simply given by 
\begin{equation}
S_\text{L}(\omega)=\frac{\Gamma}{2}|s_1(\infty)|^2 \delta(\omega-k)+\frac{\Gamma}{2\pi} \re I_3(\omega).
\end{equation}
Note that number conservation is indeed guaranteed,
\begin{equation}
\langle a^\dagger_\text{R}a_\text{R}\rangle+\langle a^\dagger_\text{L}a_\text{L}\rangle
=\int d\omega\left(S_\text{R}(\omega)+S_\text{L}(\omega)\right)=A^2.
\end{equation}

Here comes the key point: If one Taylor-expands the incoherent power spectrum in terms of the driving amplitude $A$, the lowest-order term is $\mathcal{O}(A^4)$, and it would be the one calculated from Eq.~\eqref{eq:power spectrum in scattering theory} multiplying by $A^4$ for dimensional reasons; that is,  
\begin{equation}
S^\text{H-L}(\omega) = S^\text{L-S}(\omega) A^4 + \mathcal{O}(A^6)+\cdots.\label{eq:H-L=L-S}
\end{equation}
Therefore, the two approaches, the H-L equation and the L-S formalism, give a consistent result in the weak driving limit. 
In other words, \textit{in the weak driving limit the power spectrum is solely contributed by two-photon scattering processes} \cite{RephaeliPRA11,KocabasPRA12}.
Note that we have also confirmed this consistency for the semi-infinite, single-qubit case considered by Ref.~\cite{KoshinoNJP12}.

\subsection{Two distant qubits}
For two qubits the Markovian approximation is needed for an analytical solution, which is also introduced in Ref.~\cite{KoshinoNJP12}. Based on the single-qubit case, here we list the matrix $D$ derived using the same procedure:
\begin{widetext}
	\setcounter{MaxMatrixCols}{15}
	\begin{equation}
	\Resize{18cm}{
		D=\begin{pmatrix}
		i\delta\omega-\frac{\Gamma}{2} & 0 & -\frac{\Gamma}{2} e^{ik_0L} & 0 & 2i\mathfrak{N} & 0 & 0 & 0 & 0 & 0 & \Gamma e^{ik_0L} & 0 & 0 & 0 & 0 \\ 
		0 & -i\delta\omega-\frac{\Gamma}{2} & 0 & -\frac{\Gamma}{2}e^{-ik_0L} & -2i\mathfrak{N}^* & 0 & 0 & 0 & 0 & 0 & 0 & \Gamma e^{-ik_0L}  & 0 & 0 & 0  \\ 
		-\frac{\Gamma}{2} e^{ik_0L} & 0 & i\delta\omega-\frac{\Gamma}{2} & 0 & 0 & 2i\mathfrak{N}^* & 0 & 0 & 0 & 0 & 0 & 0 & \Gamma e^{ik_0L} & 0 & 0  \\ 
		0 & -\frac{\Gamma}{2} e^{-ik_0L} & 0 &  -i\delta\omega-\frac{\Gamma}{2} & 0 & -2i\mathfrak{N} & 0 & 0 & 0 & 0 & 0 & 0 & 0 & \Gamma e^{-ik_0L} & 0  \\ 
		i\mathfrak{N}^* & -i\mathfrak{N} & 0 & 0 & -\Gamma & 0 & -\frac{\Gamma}{2} e^{ik_0L}  & -\frac{\Gamma}{2} e^{-ik_0L}  & 0 & 0 & 0 & 0 & 0 & 0 & 0  \\
		0 & 0 & i\mathfrak{N} & -i\mathfrak{N}^*  & 0 & -\Gamma & -\frac{\Gamma}{2} e^{-ik_0L}  & -\frac{\Gamma}{2} e^{ik_0L} & 0 & 0 & 0 & 0 & 0 & 0 & 0  \\ 
		0 & -i \mathfrak{N}^*  & i \mathfrak{N}^*  & 0 & -\frac{\Gamma}{2} e^{ik_0L} & -\frac{\Gamma}{2} e^{-ik_0L} & -\Gamma & 0 & 0 & 0 & -2i\mathfrak{N}^*  & 0 & 0 & 2i\mathfrak{N}^*  & 2\Gamma\cos k_0L  \\
		i\mathfrak{N} & 0 & 0 & -i\mathfrak{N}  & -\frac{\Gamma}{2} e^{-ik_0L} & -\frac{\Gamma}{2} e^{ik_0L} & 0 & -\Gamma & 0 & 0 & 0 & 2i\mathfrak{N}  & -2i\mathfrak{N}  & 0 & 2\Gamma\cos k_0L  \\ 
		-i\mathfrak{N}^*  & 0 & -i\mathfrak{N}  & 0 & 0 & 0 & 0 & 0 & 2i\delta\omega-\Gamma & 0 & 2i\mathfrak{N}  & 0 & 2i\mathfrak{N}^*  & 0 & 0  \\
		0 & i\mathfrak{N}  & 0 & i\mathfrak{N}^*  & 0 & 0 & 0 & 0 & 0 & -2i\delta\omega-\Gamma & 0 & -2i\mathfrak{N}^* & 0 & -2i\mathfrak{N} & 0  \\ 
		0 & 0 & 0 & 0 & -i\mathfrak{N}^* & 0 & -i\mathfrak{N} & 0 & i\mathfrak{N}^* & 0 & i\delta\omega-\frac{3\Gamma}{2} & 0 & -\frac{\Gamma}{2} e^{-ik_0L} & 0 & 2i\mathfrak{N}^*  \\
		0 & 0 & 0 & 0 & i\mathfrak{N} & 0 & 0 & i\mathfrak{N}^* & 0 & -i\mathfrak{N} & 0 & -i\delta\omega-\frac{3\Gamma}{2} & 0 & -\frac{\Gamma}{2} e^{ik_0L} & -2i\mathfrak{N}  \\
		0 & 0 & 0 & 0 & 0 & -i\mathfrak{N} & 0 & -i\mathfrak{N}^* & i\mathfrak{N} & 0 & -\frac{\Gamma}{2} e^{-ik_0L} & 0 & i\delta\omega-\frac{3\Gamma}{2} & 0 & 2i\mathfrak{N}  \\
		0 & 0 & 0 & 0 & 0 & i\mathfrak{N}^* & i\mathfrak{N}  & 0 & 0 & -i\mathfrak{N}^*  & 0 & -\frac{\Gamma}{2}e^{ik_0L} & 0 & -i\delta\omega-\frac{3\Gamma}{2} & -2i\mathfrak{N}^*   \\ 
		0 & 0 & 0 & 0 & 0 & 0 & 0 & 0 & 0 & 0 & i\mathfrak{N} & -i\mathfrak{N}^*  & i\mathfrak{N}^* & -i\mathfrak{N}  & -2\Gamma
		\end{pmatrix},}
	\end{equation}
	where $\mathfrak{N}=\sqrt{\Gamma/2}Ae^{-ik_0L/2}$. For equal-time correlations, the vectors $\mathbf{S}$ and $\mathbf{F}$ are given by
	\setcounter{MaxMatrixCols}{15}
	\begin{equation}
	\mathbf{S}(t)=
	\begin{pmatrix} s_1(t) \\ s_1(t)^* \\ s_2(t) \\ s_2(t)^* \\ s_3(t) \\ s_4(t) \\ s_5(t) \\ s_5(t)^* \\ s_6(t) \\ s_6(t)^* \\ s_7(t) \\ s_7(t)^* \\ s_8(t) \\ s_8(t)^* \\ s_9(t)
	\end{pmatrix}
	=\begin{pmatrix} 
	\langle\sigma_{1-}(t)\rangle e^{ikt} \\ \langle\sigma_{1+}(t)\rangle e^{-ikt} \\ \langle\sigma_{2-}(t)\rangle e^{ikt} \\ \langle\sigma_{2+}(t)\rangle e^{-ikt} \\ \langle\sigma_{1+}(t)\sigma_{1-}(t)\rangle  \\
	\langle\sigma_{2+}(t)\sigma_{2-}(t)\rangle  \\ \langle\sigma_{1+}(t)\sigma_{2-}(t)\rangle  \\ \langle\sigma_{2+}(t)\sigma_{1-}(t)\rangle  \\ \langle\sigma_{1-}(t)\sigma_{2-}(t)\rangle e^{2ikt}  \\
	\langle\sigma_{1+}(t)\sigma_{2+}(t)\rangle e^{-2ikt}  \\
	\langle\sigma_{1+}(t)\sigma_{1-}(t)\sigma_{2-}(t)\rangle e^{ikt}  \\
	\langle\sigma_{1+}(t)\sigma_{1-}(t)\sigma_{2+}(t)\rangle e^{-ikt}  \\
	\langle\sigma_{2+}(t)\sigma_{2-}(t)\sigma_{1-}(t)\rangle e^{ikt}  \\
	\langle\sigma_{2+}(t)\sigma_{2-}(t)\sigma_{1+}(t)\rangle e^{-ikt}  \\
	\langle\sigma_{1+}(t)\sigma_{1-}(t)\sigma_{2+}(t)\sigma_{2-}(t)\rangle
	\end{pmatrix},\quad
	\mathbf{F}=
	\begin{pmatrix} -i\mathfrak{N} \\ i\mathfrak{N}^* \\ -i\mathfrak{N}^* \\ i\mathfrak{N} \\ 0 \\0\\0\\0 \\0\\0\\0 \\0\\0\\0 \\0
	\end{pmatrix}.
	\end{equation}
	Following the same procedure, one can construct the two- and multiple-time correlations using the same matrix $D$. We checked that the results agree with those of Ref.~\cite{LalumierePRA13}, as well as those obtained from the L-S formalism presented in the main text. Finally, we note that for $N$ qubits in the Markovian regime, the size of the matrix $D$ grows as $(4^N-1)^2$, where the factor of 4 arises because each qubit can be represented by $\sigma_{x, y, z}$ or $1$ and subtracting one means not to consider the trivial correlation $\langle 1^N \rangle$.

	\section{Two-photon Transmission Probability}
	\label{appendix:Two-photon Transmission Probability}
	
	In this appendix we consider the transmission and reflection probabilities using the L-S formalism. Specifically, we calculate the expectation values $\langle \psi_2 | a^\dagger_\text{R}a_\text{R}|\psi_2\rangle$ and $\langle \psi_2 | a^\dagger_\text{L}a_\text{L}|\psi_2\rangle$. First of all, this calculation serves as a consistency check, and it can also be compared with the H-L approach. Secondly, it justifies our argument that it is not necessary to consider the transmission $g_2$ when driving the system resonantly. Here we focus on two illustrative examples: a single qubit coupled to both infinite and semi-infinite waveguides.
	
	We start from the transmission probability for the infinite waveguide case. One first inserts the one-particle identity operator $\mathcal{I}_1$ and employs Eq.~\eqref{eq:two-photon matrix element},
	\begin{align}
	\langle \psi_2 | a^\dagger_\text{R}(x_0)a_\text{R}(x_0)|\psi_2\rangle
	&= \int dk\biggl(|_\text{R}\langle\phi_1(k) | a_\text{R}(x_0) | \psi_2(k_1, k_2)\rangle|^2
	+|_\text{L}\langle\phi_1(k) | a_\text{R}(x_0) | \psi_2(k_1, k_2)\rangle|^2\biggr)\nonumber\\
	&=\int dk \biggl|\left(\frac{e^{ik_1x_0}}{\sqrt{4\pi}}t(k_1)\delta(k-k_2)
	+k_1\leftrightarrow k_2\right)
	-RR(k, x_0)G^{-1}e(k_1)e(k_2)\biggr|^2\nonumber\\
	&\quad\quad+\int dk\biggl|LR(k, x_0)G^{-1}e(k_1)e(k_2)\biggr|^2 . 
	\end{align}
	One can see that the above expression contains three parts: the plane waves, the bound state, and their interference. The integration over the plane waves is
	\begin{equation}
	\int dk \biggl|\frac{e^{ik_1x_0}}{\sqrt{4\pi}}t(k_1)\delta(k-k_2)
	+k_1\leftrightarrow k_2\biggr|^2
	=\left(\frac{|t(k_1)|^2}{4\pi}\delta(0)+\frac{t(k_1)^*t(k_2)}{4\pi}\delta(k_1-k_2)\right)+k_1\leftrightarrow k_2,
	\end{equation}
	and the interference term is 
	\begin{equation}
	-\int dk \left[\left(\frac{e^{ik_1x_0}}{\sqrt{4\pi}}t(k_1)\delta(k-k_2)
	+k_1\leftrightarrow k_2\right)^*\times
	RR(k, x_0)G^{-1}e(k_1)e(k_2)\right]-\textrm{h.c.}=0.
	\end{equation}
	Finally, the bound state term is 
	\begin{equation}
	\int dk\left[\bigl|RR(k, x_0)G^{-1}e(k_1)e(k_2)\bigr|^2+\bigl|LR(k, x_0)G^{-1}e(k_1)e(k_2)\bigr|^2\right]=\frac{2\Gamma^3/\pi^2}{\left[(E-2\omega)^2+\Gamma^2\right]^2},
	\end{equation}
\end{widetext}
which could also be derived by integrating Eq.~\eqref{eq:single qubit S(w) infinite waveguide} over $\omega$ and dividing by $2\pi$. Combining the three pieces together and taking $k_1=k_2=E/2$, we have
\begin{equation}
\langle  \psi_2 | a^\dagger_\text{R}a_\text{R}|\psi_2\rangle = |t(E/2)|^2\frac{\delta(0)}{\pi}+\frac{2\Gamma^3/\pi^2}{\left[(E-2\omega)^2+\Gamma^2\right]^2}, \label{eq:two-photon transmission infinite}
\end{equation}
which is indeed independent of $x_0$ as it should.

The presence of the infinity $\delta(k=0)$, proportional to the ``volume'' of the system, in Eq.~\eqref{eq:two-photon transmission infinite} seems rather awkward and needs interpretation. 
Our one-photon input state, implicitly defined in Eq.~\eqref{eq:single particle eigenstate semi-infinite}, is
\begin{equation}
|k\rangle=\int dx \frac{e^{ikx}}{\sqrt{2\pi}}a^\dagger_\text{R}(x)|0\rangle,
\end{equation}
which gives $\langle k|k'\rangle=\delta(k-k')$ and $\langle k|a_\text{R}^\dagger a_\text{R}|k\rangle=1/2\pi$. Using this to construct the two-photon input state, $|k_1, k_2\rangle=|k_1\rangle\otimes|k_2\rangle/\sqrt{2}$, one obtains \begin{equation}
\langle k_1, k_2|a_\text{R}^\dagger a_\text{R}|k_1, k_2\rangle=2\times\frac{\delta(0)}{2\pi}
\end{equation}
for $k_1=k_2$. This is dimensionally correct and tells us that the ``number of photons'' injected into the system is $\delta(0)/\pi$. Therefore, dividing Eq.~\eqref{eq:two-photon transmission infinite} by $\delta(0)/\pi$ gives the two-photon transmission probability
\begin{eqnarray}
\text{T}_2 &=& \frac{\langle  \psi_2 | a^\dagger_\text{R}a_\text{R}|\psi_2\rangle}{\delta(0)/\pi} \nonumber \\
&=& |t(E/2)|^2+\frac{4\Gamma^3}{\left[(E-2\omega)^2+\Gamma^2\right]^2}\frac{1}{2\pi\delta(0)}.
\end{eqnarray}
This dimensionless expression is transparent: the first term is the single-photon transmission probability $\text{T}=|t(k)|^2$, and the second term is the two-photon correction. Compare this expression with that calculated from the H-L equation (cf. Appendix~\ref{appendix:input-output theory one-qubit}); in the weak driving limit, the results agree upon requiring $A^2=1/2\pi\delta(0)$. Therefore, if the driving is weak enough ($A^2\ll\Gamma$), the two-photon correction to the transmission probability is negligible and we do not have to consider the transmission $g_2$ when driving resonantly (so $\text{T}=0$). 

One could do the calculation of $\langle  \psi_2 | a^\dagger_\text{L}a_\text{L}|\psi_2\rangle$ in a similar way. The three pieces (plane wave, interference, bound state) are 
\begin{equation}
|r(E/2)|^2\frac{\delta(0)}{\pi},\; \frac{-4\Gamma^3/\pi^2}{\left[(E-2\omega)^2+\Gamma^2\right]^2}, \;
\frac{2\Gamma^3/\pi^2}{\left[(E-2\omega)^2+\Gamma^2\right]^2},
\end{equation}
respectively. We note that in contrast to the transmission, the interference term in reflection is not zero and that the last term (bound state) is the same as that for transmission. Therefore, the two-photon reflection probability is given by
\begin{eqnarray}
\text{R}_2 &=& \frac{\langle  \psi_2 | a^\dagger_\text{L}a_\text{L}|\psi_2\rangle}{\delta(0)/\pi} \nonumber \\
&=& |r(E/2)|^2-\frac{4\Gamma^3}{\left[(E-2\omega)^2+\Gamma^2\right]^2}\frac{1}{2\pi\delta(0)}.
\end{eqnarray}
One can see that indeed $\text{T}_2+\text{R}_2=\text{T}+\text{R}=1$ holds, which relies on the precise interference between the plane wave and bound state parts.

This precise interference can also be seen in the semi-infinite waveguide case. Without transmission, one only needs to compute $\langle  \psi_2 | a^\dagger_\text{L}a_\text{L}|\psi_2\rangle$ and the three terms are given by (in the Markovian regime)
\begin{equation}
|r(E/2)|^2\frac{\delta(0)}{\pi},\; \frac{-16\tilde{\Gamma}^3/\pi^2}{\left[(E-2\tilde{\omega})^2+\tilde{\Gamma}^2\right]^2},\;
\frac{16\tilde{\Gamma}^3/\pi^2}{\left[(E-2\tilde{\omega})^2+\tilde{\Gamma}^2\right]^2}.
\end{equation}
As a result, the last two terms cancel exactly, leaving $\text{R}_2=1$. This completes the consistency check of our theory.

\bibliography{WQED_2015,\jobname}

\begin{thebibliography}{113}
\bibitem [{pre()}]{precision}%
  \BibitemOpen
  \href@noop {} {}\bibinfo {note} {Though we give the incoming photon
  frequencies with an accuracy of $0.01\Gamma$ in
  Figs.~\ref{fig:PSD_Piover2_infinite} and \ref{fig:PSD_Piover4_infinite},
  considerably greater precision is needed in doing the calculation because of
  extreme sensitivity to the driving frequency as $N$ increases.}\BibitemShut
  {Stop}%
\end{thebibliography}
\end{document}